\documentclass[11pt]{article}
\usepackage[margin=1in]{geometry}
\usepackage[T1]{fontenc}
\usepackage{lmodern}
\usepackage{amsmath,amssymb}
\usepackage{graphicx}
\usepackage{multirow}
\usepackage{rotating}
\usepackage{url}
\usepackage[authoryear,round]{natbib}
\usepackage[final,colorlinks=true,linkcolor=blue,citecolor=blue,urlcolor=blue]{hyperref}
\usepackage[sort&compress,nameinlink]{cleveref}
\usepackage{microtype}
\usepackage{titlesec}
\usepackage{fancyhdr}

\makeatletter
\newcommand{\authors}[1]{\gdef\arxivauthors{#1}}
\newcommand{\affil}[1]{\textsuperscript{#1}}
\newcommand{\affiliation}[2]{\gdef\arxivaffiliation{\textsuperscript{#1}#2}}
\newcommand{\correspondingauthor}[2]{\gdef\arxivcorresponding{Corresponding author: #1 (\href{mailto:#2}{#2})}}
\newcommand{\printarxivtitle}{%
  \begin{center}
    {\LARGE\bfseries \@title\par}
    \vspace{1em}
    {\large \arxivauthors\par}
    \vspace{0.5em}
    {\small \arxivaffiliation\par}
    \vspace{0.5em}
    {\small \arxivcorresponding\par}
  \end{center}
  \vspace{1em}
}
\makeatother
\newenvironment{keypoints}{\printarxivtitle\section*{Key Points}\begin{itemize}}{\end{itemize}}
\newcommand{\acknowledgments}{\section*{Acknowledgments}}

\titleformat{\paragraph}[runin]
  {\normalfont\normalsize\bfseries}{\theparagraph}{0.75em}{}
\titlespacing*{\paragraph}
  {\parindent}{1.5ex plus .3ex}{0.75em}
\titleformat{\subparagraph}[runin]
  {\normalfont\normalsize\bfseries}{\thesubparagraph}{0.75em}{}
\titlespacing*{\subparagraph}
  {2.5\parindent}{1.25ex plus .2ex}{0.75em}

\crefformat{equation}{(eq~#2#1#3)}
\crefrangeformat{equation}{(eqs~#3#1#4--#5#2#6)}
\crefmultiformat{equation}{(eqs~#2#1#3}{, #2#1#3)}{, #2#1#3}{, #2#1#3)}
\crefrangemultiformat{equation}{(eqs~#3#1#4--#5#2#6}{, #3#1#4--#5#2#6}{, #3#1#4--#5#2#6}{, #3#1#4--#5#2#6)}
\crefformat{figure}{(fig~#2#1#3)}
\crefrangeformat{figure}{(figs~#3#1#4--#5#2#6)}
\crefmultiformat{figure}{(figs~#2#1#3}{, #2#1#3)}{, #2#1#3}{, #2#1#3)}
\crefrangemultiformat{figure}{(figs~#3#1#4--#5#2#6}{, #3#1#4--#5#2#6}{, #3#1#4--#5#2#6}{, #3#1#4--#5#2#6)}
\crefformat{table}{(tab~#2#1#3)}
\crefrangeformat{table}{(tabs~#3#1#4--#5#2#6)}
\crefmultiformat{table}{(tabs~#2#1#3}{, #2#1#3)}{, #2#1#3}{, #2#1#3)}
\crefrangemultiformat{table}{(tabs~#3#1#4--#5#2#6}{, #3#1#4--#5#2#6}{, #3#1#4--#5#2#6}{, #3#1#4--#5#2#6)}
\crefformat{section}{(sec~#2#1#3)}
\crefrangeformat{section}{(secs~#3#1#4--#5#2#6)}
\crefmultiformat{section}{(secs~#2#1#3}{, #2#1#3)}{, #2#1#3}{, #2#1#3)}
\crefrangemultiformat{section}{(secs~#3#1#4--#5#2#6}{, #3#1#4--#5#2#6}{, #3#1#4--#5#2#6}{, #3#1#4--#5#2#6)}
\crefformat{subsection}{(sec~#2#1#3)}
\crefrangeformat{subsection}{(secs~#3#1#4--#5#2#6)}
\crefmultiformat{subsection}{(secs~#2#1#3}{, #2#1#3)}{, #2#1#3}{, #2#1#3)}
\crefrangemultiformat{subsection}{(secs~#3#1#4--#5#2#6}{, #3#1#4--#5#2#6}{, #3#1#4--#5#2#6}{, #3#1#4--#5#2#6)}
\crefformat{subsubsection}{(sec~#2#1#3)}
\crefrangeformat{subsubsection}{(secs~#3#1#4--#5#2#6)}
\crefmultiformat{subsubsection}{(secs~#2#1#3}{, #2#1#3)}{, #2#1#3}{, #2#1#3)}
\crefrangemultiformat{subsubsection}{(secs~#3#1#4--#5#2#6}{, #3#1#4--#5#2#6}{, #3#1#4--#5#2#6}{, #3#1#4--#5#2#6)}
\crefformat{paragraph}{(sec~#2#1#3)}
\crefrangeformat{paragraph}{(secs~#3#1#4--#5#2#6)}
\crefmultiformat{paragraph}{(secs~#2#1#3}{, #2#1#3)}{, #2#1#3}{, #2#1#3)}
\crefrangemultiformat{paragraph}{(secs~#3#1#4--#5#2#6}{, #3#1#4--#5#2#6}{, #3#1#4--#5#2#6}{, #3#1#4--#5#2#6)}
\crefformat{subparagraph}{(sec~#2#1#3)}
\crefrangeformat{subparagraph}{(secs~#3#1#4--#5#2#6)}
\crefmultiformat{subparagraph}{(secs~#2#1#3}{, #2#1#3)}{, #2#1#3}{, #2#1#3)}
\crefrangemultiformat{subparagraph}{(secs~#3#1#4--#5#2#6}{, #3#1#4--#5#2#6}{, #3#1#4--#5#2#6}{, #3#1#4--#5#2#6)}
\crefformat{appendix}{(app~#2#1#3)}
\crefrangeformat{appendix}{(apps~#3#1#4--#5#2#6)}
\crefmultiformat{appendix}{(apps~#2#1#3}{, #2#1#3)}{, #2#1#3}{, #2#1#3)}
\crefrangemultiformat{appendix}{(apps~#3#1#4--#5#2#6}{, #3#1#4--#5#2#6}{, #3#1#4--#5#2#6}{, #3#1#4--#5#2#6)}

\begin{document}

\newcommand{\symup}[1]{\mathrm{#1}}
\newcommand{\symscr}[1]{\mathcal{#1}}
\newcommand{\symbfup}[1]{\mathbf{#1}}
\newcommand{\symbfit}[1]{\mathbf{#1}}
\renewcommand{\autoref}[1]{\cref{#1}}

\newcommand{\ddt}[1]{\frac{\symup{d} #1}{\symup{d} t}}
\newcommand{\pdt}[1]{\frac{\partial #1}{\partial t}}
\newcommand{\nab}{\symbfup{\nabla}}
\renewcommand{\vec}[1]{\symbfit{#1}}

\title{Elastic Non-Uniform FFT (ENUFFT)\\ spectral reconstruction of irregularly sampled orography on unstructured grids}

\authors{Tridib Banerjee\affil{1}, Felix Jochum\affil{1}, Ulrich Achatz\affil{1}}

\affiliation{1}{Institute for Atmospheric and Environmental Sciences, Goethe University Frankfurt, Frankfurt, Germany}

\correspondingauthor{Tridib Banerjee}{banerjee@iau.uni-frankfurt.de}

\begin{keypoints}
\item The new method recovers local terrain spectra directly from irregularly sampled data on non-rectangular unstructured model grids
\item The new adaptive mode selection reduces retained spectral modes while still preserving flow-dependent launch power
\item Tests show accuracy comparable to, and energy conservation substantially better than existing methods
\end{keypoints}

\begin{abstract}
Subgrid-scale orography remains a leading source of uncertainty in numerical modeling because terrain spectra must be recovered from irregularly sampled elevation data and then reduced to a flow-dependent launch budget for parameterizations. Existing approaches are limited either by assuming regular samples on rectangular grids or by fitting coefficients whose truncation and regularization effects become embedded in the spectrum. None achieves dynamic, flow-dependent truncation. This study introduces an Elastic Non-Uniform Fast Fourier Transform (ENUFFT) framework that computes local Fourier coefficients directly from irregularly sampled orography on unstructured grids, without interpolation or fitting. It combines a type-1 NUFFT with local windowing, quadrature weights, and a new Elastic Mode Selection (EMS) algorithm for retaining a local flow-dependent subset of modes. ENUFFT is compared with the strongest relevant existing method in a monochromatic and a real Alpine terrain test. In both cases, it recovers peak amplitude and direction comparably while significantly compacting the spectra (monochromatic $\sim25\,\%$, Alpine $\sim60\,\%$). It also satisfies the Parseval condition more closely with its spectral variance (energy) deviating from reference by $\sim14$--$24\,\%$ versus $\sim500$--$122{,}000\,\%$ for the existing method. Its EMS is additionally tested in a mountain-wave test where it reduces the launch spectrum by $\ge75\,\%$ while keeping launch-power loss $\leq7\,\%$. Along with better compute scaling, ENUFFT is thus a computationally efficient, physically interpretable framework that can make Fourier-based orographic source descriptions practical for spectral-budget-aware parameterizations.
\end{abstract}

\section*{Plain Language Summary}
In weather and climate modeling, flow over the earth's surface is crucial, and representing it poorly causes large forecasting errors. At the heart of this decades-old problem lie two facts. First, the surface is more complex than any model can ever resolve. Second, the part of them that matters changes with the flow. So a workable model must keep only the features most relevant to the flow, and do so dynamically. The best approaches use a tool called Fourier analysis that breaks complex patterns into simpler ones, but it assumes the surface is regularly sampled and evenly boxed, which real terrain is not. Past methods either oversimplified the surface or approximated it by fitting. Flow-dependent filtering was never achieved.

This study develops a new method, the Elastic Non-Uniform Fast Fourier Transform (ENUFFT), that extracts those patterns directly from irregular terrain and keeps the ones that matter most for each flow. Tested on synthetic and real terrain, it matches today's best methods while adding dynamic filtering, large computational savings, and better energy conservation. Overall, it is a powerful new tool for representing the earth's surface in models, one that can lead to more accurate forecasts.

\section{Introduction}\label{sec:introduction}
The surface of earth varies over a continuous range of spatial scales, from continental relief down to meter-scale roughness, and across most of that range its statistics remain mostly scale invariant \citep{gagnon2006multifractal,balmino1993spectra}. A spectral description is a standard way to characterize such structure, since it orders topography by wavelength and orientation,  exposing the characteristic scales, anisotropy, and roughness that bulk descriptors would otherwise hide \citep{perron2008spectral,booth2009automated}. Such description matters most when model grids cannot physically resolve the scales of interest.

Terrain spectra are most needed in the parameterization of subgrid-scale orography. Schemes for orographic gravity-wave drag were first introduced decades ago to counter the systematic circulation biases that unresolved mountains would otherwise produce \citep{palmer1986alleviation,mcfarlane1987effect}. Later schemes then added anisotropy, blocking, form drag, and directional dependence to build the mature family now in operational use \citep{lott1997subgrid,gregory1998gravity,scinocca2000drag,webster2003representation,beljaars2004formdrag,choi2015updated,xie2021implementation}. Even so, reviews still single out the orographic source as one of the least constrained elements of these schemes \citep{kim2003overview,teixeira2014physics,plougonven2020guide} despite parameterized orographic drag being known to exert a leading-order influence on the simulated circulation while carrying substantial structural uncertainty \citep{sandu2016impacts,hajkova2024parameterized}. The same demand arises well beyond drag closures, for example in the scale analysis of kinetic energy and dissipation on unstructured ocean and atmospheric meshes \citep{juricke2023scaleanalysis} and in the estimation of spectra from scattered particle samples in meshless and Lagrangian solvers \citep{shi2013interpolation}.

Meeting this need in practice is complicated by the available data and grids. Surface elevation comes as digital elevation models (DEMs) whose resolution varies widely, with near-global coverage at $30\,\mathrm{m}$ \citep{copernicusdataspace2026,jaxaaw3d30v4} alongside national and regional lidar products at $1$ to $2\,\mathrm{m}$ \citep{usgs_s1m2026,swisstopo2026} and polar mosaics at $2\,\mathrm{m}$ \citep{arcticdem2026,rema2026}, a constraint that matters because spectral fidelity depends on both resolution \citep{elvidge2019uncertainty} and sampling density \cref{app:discrete_spectral_fidelity}. The grids are no more convenient. Global models increasingly replace the latitude-longitude mesh with quasi-uniform icosahedral and other unstructured grids \citep{staniforth2012horizontal,Zangl_2015}, so the local spectral window is a projected, rotated, and clipped model cell rather than an axis-aligned rectangle. Even when the source DEM is a regular raster, spectral analysis on a model cell is still not a regular-grid problem, as seen in operational ICON preprocessing, which aggregates raw \citet{globe1999}, \citet{astergdem2019}, and MERIT \citep{yamazaki2017merit} topography onto triangular cells and screens it with land-sea information \citep{extparfortran2026,dwdicondatabase2026}.

However, all existing methods of obtaining terrain spectra each fall short of what irregularly sampled, unstructured data demands. The fast Fourier transform that makes spectral analysis practical assumes equally spaced samples on a rectangular domain and does not carry over to scattered points on an unstructured cell \citep{cooley1965analgorithmf}. In cases where a high-resolution raster patch was available and the spectra were obtained (either through analytic base-flux closures \citep{garner2005closure} or scale-aware and directionally resolved drag schemes \citep{vanniekerk2021scaleaware,vanniekerk2023partialcritical}) and were successfully used to reduce precipitation and near-surface forecast biases \citep{liu2023improved}, all such formulations still presumed a regular raster patch in each cell. For genuinely unstructured grids the closest method is the Constrained Spectral Approximation of \citet{chew2024aconstrained}, which fits coefficients by regularized least squares, yet fitting embeds truncation and regularization directly into the coefficients and can return nonphysical peaks. What has been missing is a way to obtain local Fourier coefficients directly from irregular samples while keeping a clear spectral interpretation. The non-uniform fast Fourier transforms (NUFFTs) were developed for exactly this purpose. They perform Fourier analysis of irregularly spaced data and are now standard in numerical analysis, signal processing \citep{dutt1993fastfouriert,greengard2004accelerating,barnett2019aparallelnon} and in gridding-based image reconstruction such as magnetic resonance imaging \citep{fessler2003nonuniformfa,Jackson1991Gridding,PipeMenon1999SDC}, yet they have not been turned to recovering local subgrid orographic spectra on non-rectangular cells.

A second requirement appears once the recovered spectrum must be used rather than stored. Propagation-aware gravity-wave schemes such as the ray tracer GROGRAT \citep{marks1995raytracing} and the transient parameterization MS-GWaM \citep{Voelker2024} follow individual wave packets, accounting for transience and horizontal propagation, and can represent orographic mean-flow forcing more accurately than the classical steady-state treatment \citep{Jochum2025}. That accuracy carries a cost that grows with the spectrum as the number of retained modes also becomes a runtime budget \citep{Boeloeni2021,Achatz2023}. The budget is complicated further because of the launch spectrum being flow dependent rather than fixed. Therefore, compressing the spectrum not only becomes unavoidable, but to remain physically defensible that compression must also be dynamic and flow dependent rather than a uniform cutoff, a capability no existing terrain-spectrum method provides.

Taken together, these gaps define what an orographic source framework ought to do. It should recover physically interpretable Fourier coefficients, operate on irregularly sampled terrain over unstructured local domains, and compress the launch spectrum dynamically while keeping the modes that matter. The present study meets these requirements with an Elastic NUFFT (ENUFFT) framework. ENUFFT computes local Fourier coefficients directly from irregular DEM samples without interpolation or fitting. A shape-aware elastic mode-selection algorithm then sets a cell and flow-dependent number of retained modes that preserves a target fraction of launch-relevant power under a prescribed budget. It turns the spectral compression into a physically guided allocation of the ray-launch budget rather than a uniform cutoff. Because the retained quantity can be drawn from spectral energy or from a weighted launch measure such as wave action, the same logic serves static terrain spectra and flow-dependent source spectra alike. The framework thus aims both to improve terrain representation on unstructured grids and to make Fourier-based source descriptions viable in propagation-aware schemes, where geometric fidelity and online cost must be balanced.


\section{Theory}
This section has three parts. The first demonstrates how to compute Fourier coefficients from irregularly sampled terrain on rectangular domains. The second extends the formulation to local triangular domains. The third develops an elastic and shape-aware mode-selection algorithm that can dynamically truncate any spectrum to achieve orders-of-magnitude compaction without sacrificing important modes, while also respecting prescribed power-fraction and mode-budget constraints.

\subsection{Rectangular Non-Uniform Fast Fourier Transform}\label{sec:nufft}

This subsection describes a framework for computing Fourier coefficients from irregularly sampled terrain on rectangular domains using a type-1 NUFFT, that is, a map from non-uniform DEM samples to coefficients on a uniform rectangular Fourier-mode grid \citep{dutt1993fastfouriert,greengard2004accelerating,barnett2019aparallelnon}.

\subsubsection{Setup}

Let a given grid cell be embedded in a local tangent-plane patch $D=\left[0,L_x\right]\times\left[0,L_y\right]$ and let $\left\{x_q,y_q,h_q\right\}_{q=1}^{Q}$ be the digital elevation model (DEM) within that patch, on whatever native (possibly non-uniform and non-quadrilateral) grid is available. The goal is to approximate $h\left(x,y\right)$ on this patch with the truncated Fourier series
\begin{equation}\label{eq:series}
  h\left(x,y\right) \approx \sum_{m=-\mathcal{M}}^{\mathcal{M}} \sum_{n=-\mathcal{N}}^{\mathcal{N}}
  \hat h_{m,n} e^{\,i\left(k_m x + \ell_n y\right)},
  \qquad
  k_m = \frac{2\pi m}{L_x},\quad
  \ell_n = \frac{2\pi n}{L_y}
\end{equation}
with realness enforced by $\hat h_{-m,-n}=\hat h_{m,n}^*$, where an asterisk indicates complex conjugation. Beginning with the continuous definition of Fourier coefficients on the local tangent-plane patch $D$, the Fourier coefficient for a wave vector $\mathbf{k}_{m,n}=\left(k_m,\ell_n\right)$ of a square-integrable doubly periodic orography field $h\left(x,y\right)$ is
\begin{equation}
  \hat h\left(k_m,\ell_n\right)
  = \frac{1}{\left|D\right|} \int_{0}^{L_x}\int_{0}^{L_y}
  h\left(x,y\right)\exp\left[-i\left(k_m x + \ell_n y\right)\right]\,\mathrm{d}y\,\mathrm{d}x,
  \qquad \left|D\right| = L_x L_y.
  \label{eq:cont_fourier}
\end{equation}
For the truncated series in \cref{eq:series}, the relevant wavenumbers are,
\begin{equation}
  k_m = \frac{2\pi m}{L_x},\qquad \ell_n = \frac{2\pi n}{L_y},
  \qquad \left|m\right|\le\mathcal{M},\ \left|n\right|\le\mathcal{N},
\end{equation}
so that one seeks $\hat h_{m,n}\approx \hat h\left(k_m,\ell_n\right)$. In practice, the DEM provides only finitely many samples $\left\{x_q,y_q,h_q\right\}_{q=1}^{Q}$ inside $D$. A standard way to approximate the integral in \cref{eq:cont_fourier} is to use numerical quadrature (i.e., a weighted sum)
\begin{equation}
  \frac{1}{\left|D\right|}\sum_{q=1}^{Q} w_q h_q
  \exp\left[-i\left(k_m x_q + \ell_n y_q\right)\right],
  \label{eq:quad}
\end{equation}
where $w_q>0$ are quadrature area weights that approximate $\mathrm{d}x\,\mathrm{d}y$. For simplicity, and because DEM sampling is typically dense and near-uniform at the scales of interest, equal weights $w_q = \frac{1}{Q}\left|D\right|$ can be used. In that case,
\begin{equation}
  \hat h_{m,n}
  \equiv \frac{1}{Q} \sum_{q=1}^{Q} h_q\,
  \exp\left[-i\left(k_m x_q + \ell_n y_q\right)\right]
  \label{eq:nufft_sum}
\end{equation}
represents a natural discrete approximation to \cref{eq:cont_fourier}. This is valid for any point distribution, uniform or not. On a strictly equidistant grid, one has
\begin{equation}
  x_q = r\Delta x,\quad y_q = s\Delta y,
  \qquad \Delta x = \frac{L_x}{N_x},\ \Delta y = \frac{L_y}{N_y},
\end{equation}
with $q$ denoting the flattened version of $\left(r,s\right)$, and \cref{eq:nufft_sum} reduces to the familiar two-dimensional discrete Fourier transform (DFT)
\begin{equation}
  \hat h_{m,n}
  = \frac{1}{N_x N_y}\sum_{r=0}^{N_x-1}\sum_{s=0}^{N_y-1}
  h_{r,s} \exp\left[-2\pi i\left(\frac{mr}{N_x} + \frac{ns}{N_y}\right)\right],
\end{equation}
which can typically be evaluated in $\mathcal{O}\left[N_x N_y \log\left(N_x N_y\right)\right]$ using FFT algorithms \citep{cooley1965analgorithmf}. The discrete sum in \cref{eq:nufft_sum} is algebraically identical to the discrete Fourier transform, except that the sample locations $\left(x_q,y_q\right)$ lie on an irregular grid instead of a regular one. The complication is entirely computational. When the sampling is irregular, the sum cannot be factorized into independent transforms in $x$ and $y$, so standard FFT algorithms no longer apply. Evaluating this sum directly for all spectral modes of which there are $\left(2\mathcal{M}+1\right)\left(2\mathcal{N}+1\right)= \mathcal{O}\left(\mathcal{M}\mathcal{N}\right)$, requires looping over all $Q$ DEM samples for each mode. The resulting brute-force cost $ \mathcal{O}\left(Q\,\mathcal{M}\mathcal{N}\right)$ quickly becomes prohibitive with even conservative grids having thousands of cells.

\subsubsection{Rectangular NUFFT}\label{sec:nufft_formulation_efficient_discrete_evaluation}

A non-uniform fast Fourier transform (NUFFT) is a family of algorithms that evaluate sums of the form \cref{eq:nufft_sum} in subquadratic time \citep{dutt1993fastfouriert,greengard2004accelerating,barnett2019aparallelnon}. The framework below uses a type-1 NUFFT, namely the non-uniform-point to uniform-frequency transform, to obtain discrete Fourier coefficients for a given orography on rectangular domains. It involves four steps.

\paragraph{\textbf{Step 1: Convolution}}
First, an equidistant auxiliary grid $\left\{x_r,y_s\right\}$ covering $D$ is introduced, usually with slight oversampling.  Each scattered DEM sample $h_q$ is then spread to nearby grid nodes using a compact smooth kernel~$\varphi$ as,
\begin{equation}
  \tilde h_{r,s}
  \approx
  \sum_{q=1}^{Q} h_q\,
  \varphi\left(x_q-x_r\right)\varphi\left(y_q-y_s\right)
  \label{eq:spreading}
\end{equation}
on a regular grid. Because $\varphi$ is supported only on a small stencil (typically a few grid spacings), each $h_q$ affects only a handful of $\tilde h_{r,s}$, keeping this step $\mathcal{O}\left(Q\right)$. The choice of $\varphi$ must balance accuracy and efficiency through three essential properties - compactness, smoothness, and spectral localization. Compactness implies $\varphi\left(\xi\right)=0$ for $\left|\xi\right|>a$, so the spreading stencil is small and efficient to compute. Smoothness gives rapid decay of the Fourier transform~$\Phi\left(k\right)$, which suppresses aliasing and stabilizes the deconvolution step, and spectral localization requires $\Phi\left(k\right)$ to remain strictly positive on the frequency band of interest, avoiding error amplification during deconvolution.

A widely used choice that satisfies these criteria is the Kaiser-Bessel (KB) kernel. In one dimension, it is
\begin{equation}
  \varphi\left(x\right)=
  \begin{cases}
    \displaystyle \frac{1}{I_0\left(\beta\right)}
    I_0\left[\gamma\sqrt{1-\left(x/a\right)^2}\right], & \left|x\right|\le a,\\[6pt]
    0, & \left|x\right|>a,
  \end{cases}
  \label{eq:kb_compact}
\end{equation}
where $I_0$ is the modified Bessel function of the first kind, $a$ is the half-width (typically $4$--$6$ grid cells), and $\left(\gamma,\beta\right)$ are the shape parameters. The traditional form in \citet{fessler2003nonuniformfa} sets
\begin{equation}\label{eq:kb_type_2}
   \gamma=\beta=2.34
\end{equation}
for the normalization $\varphi\left(0\right)=1$, whereas a version more commonly found in libraries such as the \textit{Michigan Image Reconstruction Toolbox} \citep{MIRT} uses
\begin{equation}\label{eq:kaisal_bessel_coeffs}
    \beta=2.34,\qquad\gamma=\pi\sqrt{\left(\frac{2a\beta}{\pi}\right)^2-0.8}.
\end{equation}
As shown later, \cref{eq:kaisal_bessel_coeffs} leads to lower deviation from the DFT. For two dimensions, the separable form $\varphi\left(x\right)\varphi\left(y\right)$ is used. The KB kernel is $C^\infty$, has a well-behaved Fourier transform, and achieves high accuracy (errors $<10^{-10}$) with moderate oversampling (about $1.25$--$2$) \citep{fessler2003nonuniformfa,barnett2019aparallelnon,barnett2021aliasingerro}. Other options, such as truncated Gaussians or B-splines, are simpler but either localize poorly in frequency (B-splines) or require significantly larger oversampling to reach similar accuracy (Gaussians) \citep{fessler2003nonuniformfa}. The KB kernel provides an optimal balance of compact support, smoothness, and spectral localization and is the default choice in modern NUFFT libraries such as \textit{FINUFFT}, \textit{NFFT}, and \textit{MIRT}. For the present offline orographic preprocessing, it offers a robust and accurate way to transfer irregular DEM data to a uniform FFT-compatible grid.

  \paragraph{\textbf{Step 2: Approximate Normalization}}
The continuous analog to \cref{eq:spreading} is given by
\begin{equation}
    \tilde h^{\mathrm{cont}}_{r,s}=\int\int h\left(u,v\right)\varphi\left(u-x_r\right)\varphi\left(v-y_s\right)\,du\,dv.
\end{equation}
To approximate this integral correctly for the $Q$ DEM samples on a discrete grid with spacing $\left(\Delta x,\Delta y\right)$, one computes
\begin{equation}\label{eq:normalize_kb_spreading}
    \tilde h_{r,s}^{\mathrm{true}}=\Delta A\sum_{q=1}^{Q} h_q
  \varphi\left(x_q-x_r\right)\varphi\left(y_q-y_s\right),
\end{equation}
where $\Delta A=L_xL_y/Q=N_xN_y\Delta x\Delta y/Q$ is the Monte Carlo quadrature approximate area weight. This weight is approximate because the DEM samples are not placed on a uniform grid, so $\Delta A$ is not exactly the local quadrature weight of each point. Instead, it is a uniform area weight that statistically approximates the area fraction represented by each scattered sample. Although not exact, $\Delta A$ is typically small and compacts the field, making it more stable during the FFT, especially for denormalized kernels such as the one used in \cref{eq:kaisal_bessel_coeffs}.

  \paragraph{\textbf{Step 3: FFT}}
  A standard two-dimensional FFT is then applied to $\tilde h_{r,s}^{\mathrm{true}}$, yielding the approximate Fourier coefficients
  \begin{equation}
    \hat{\tilde h}_{m,n} = \mathrm{FFT}\left(\tilde h_{r,s}^{true}\right)
  \end{equation}
  on a uniform-frequency grid associated with the auxiliary grid spacings. The computational cost of this step is $\mathcal{O}\left[N_x N_y \log\left(N_x N_y\right)\right]$.

  \paragraph{\textbf{Step 4: Deconvolution}}
  Spreading with $\varphi$ corresponds, in spectral space, to multiplying the exact coefficients $\hat h_{m,n}$ by the Fourier transform of the kernel, say $\Phi\left(k_m\right)\Phi\left(\ell_n\right)$ (convolution theorem). An approximation of the desired $\hat h_{m,n}$ is thus recovered by dividing out this factor as
  \begin{equation}
    \hat h_{m,n} = \frac{1}{\Phi\left(k_m\right)\Phi\left(\ell_n\right)}\hat{\tilde h}_{m,n}.
    \label{eq:deconv}
  \end{equation}

\begin{figure}[!t]
\centering\includegraphics[width=\textwidth]{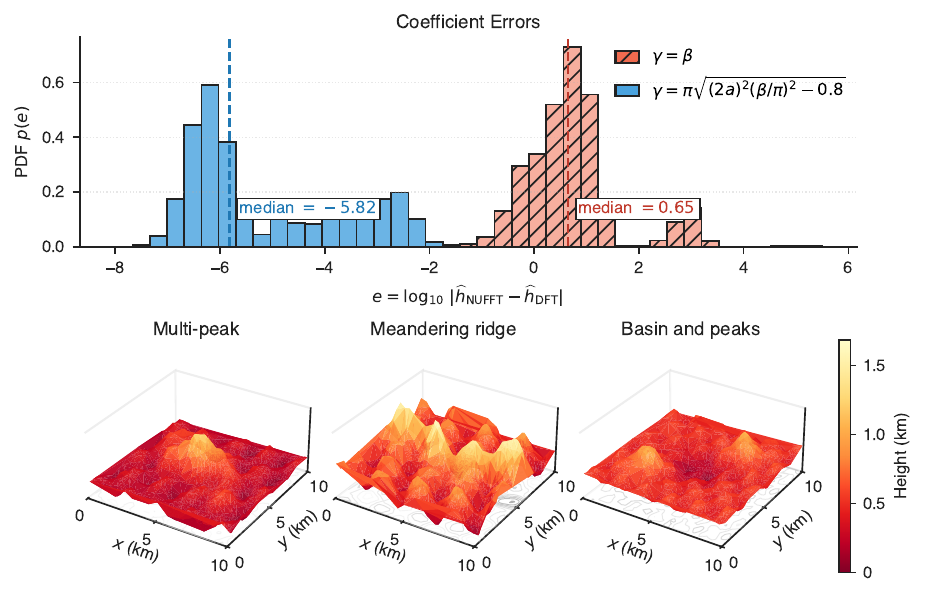}
\caption{Comparison of the baseline Kaiser-Bessel kernel in \cref{eq:kb_type_2} to the modified form in \cref{eq:kaisal_bessel_coeffs} for the NUFFT framework of \cref{sec:nufft}. Upper panel shows the pooled probability density of coefficient errors $e=\log_{10}\left|\hat h_{\mathrm{NUFFT}}-\hat h_{\mathrm{DFT}}\right|$, while lower panels show the three synthetic terrains used. Dashed vertical lines mark the median error for each kernel.}
\label{fig:nufft_kernel_comp}
\end{figure}

Through appropriate choices of grid oversampling, kernel width and kernel shape, the composite operation \cref{eq:spreading}--\cref{eq:deconv} approximates the exact sum \cref{eq:nufft_sum} within a controllable error tolerance, while essentially reducing the computational cost to
\begin{equation}\label{eq:enufft_complexity}
  \mathcal{O}\left[Q + N_x N_y \log\left(N_x N_y\right)\right]
\end{equation}
where $\left(N_x,N_y\right)$ is the number of points on the auxiliary grid. The forward transform is still given by \cref{eq:nufft_sum}, but its evaluation is accelerated by gridding and FFT algorithms. Because the orography $h\left(x,y\right)$ is static in time, the NUFFT computation of $\hat h_{m,n}$ is performed once per cell in an offline preprocessing step. At runtime, only the stored coefficients of the selected modes are needed. For the standard normalized Kaiser-Bessel pair, the Fourier transform is often written in a compact analytic form \citep{kaiser1980ontheuseofth,barnett2021aliasingerro}. In the present notation, where the denominator is $I_0\left(\beta\right)$ and the inner shape parameter is allowed to be $\gamma$, the corresponding transform can be written for real $k$ as \citep{potts2021continuouswindow},
\begin{equation}\label{eq:kb_fourier}
\Phi\left(k\right)
= \frac{2a}{I_0\left(\beta\right)}
\begin{cases}
\frac{\sinh\left[\sqrt{\gamma^2 - \left(a k\right)^2}\right]}
{\sqrt{\gamma^2 - \left(a k\right)^2}}, & \left|a k\right| < \gamma,\\
1, & \left|a k\right| = \gamma,\\
\frac{\sin\left[\sqrt{\left(a k\right)^2 - \gamma^2}\right]}
{\sqrt{\left(a k\right)^2 - \gamma^2}}, & \left|a k\right| > \gamma.
\end{cases}
\end{equation}
This is equivalent to the compact expression
\begin{equation}
\Phi\left(k\right)
= \frac{2a}{I_0\left(\beta\right)}
\frac{\sinh\left[\sqrt{\gamma^2 - \left(a k\right)^2}\right]}
{\sqrt{\gamma^2 - \left(a k\right)^2}}
\end{equation}
by analytic continuation, using $\sinh\left(i\xi\right)/\left(i\xi\right)=\sin\left(\xi\right)/\xi$. Because \cref{eq:kaisal_bessel_coeffs} is used in practice whereas only \cref{eq:kb_type_2} is commonly stated in the literature, \cref{fig:nufft_kernel_comp} provides a brief comparison of the two kernels. A detailed kernel study would exceed the scope of this work but both kernels were evaluated on three synthetic irregular terrains. Their recovered NUFFT coefficients were then compared against DFT. As illustrated, the pooled median recovery error for the optimized kernel \cref{eq:kaisal_bessel_coeffs} is orders of magnitude smaller than that for the baseline version \cref{eq:kb_type_2}\footnote{Three synthetic terrains (multi-peak, meandering ridge, and basin and peaks) were used on a rectangular $10\times10$ km domain sampled by $2000$ DEM points. The comparison retains Fourier modes $\left|m\right|,\left|n\right|\le 20$, uses $N^{\mathrm{aux}}_{x,y}=2\sigma N^{\mathrm{mode}}_{\max}$ with $\sigma=1.5$, and uses a Kaiser-Bessel kernel half-width $a=4$ grid cells. In each case, the NUFFT coefficients are compared to DFT evaluated on the same DEM points.}.

\subsubsection{Triangular Extension}\label{sec:triangular_extension}
The above formulation already permits Fourier transforms on irregularly sampled terrain without interpolation to a regular grid. However, like existing related methods, it assumes that the local domain $D$ is rectangular. This subsection extends the rectangular NUFFT framework developed in \cref{sec:nufft_formulation_efficient_discrete_evaluation} to local triangular grids. Such an extension is motivated by climate models such as ICON \citep{Zangl_2015}, which discretize the sphere using icosahedral-triangular grids rather than latitude-longitude or cubed-sphere grids. In these models, orographic spectra must be localized on each triangular cell using DEM that may be available on a finer, possibly irregular grid.

Consider a global domain tiled by a coarse triangular model grid $\mathcal{T}=\allowbreak\left\{T^{\left(1\right)},\allowbreak T^{\left(2\right)},\allowbreak\ldots,\allowbreak T^{\left(N_T\right)}\right\}$. Within each triangle $T^{\left(j\right)}$, high-resolution topographic data is available from a DEM sampled at $Q^{\left(j\right)}$ scattered points $\left\{x_q,y_q,h_q\right\}_{q=1}^{Q^{\left(j\right)}}$, where $\left(x_q,y_q\right)$ are local tangent-plane coordinates and $h_q=h\left(x_q,y_q\right)$ is the surface elevation. The DEM resolution is typically much finer than the model grid, with $Q^{\left(j\right)}\gg 1$ points potentially falling within or near each coarse triangle. The objective is to compute, for each triangle $T^{\left(j\right)}$, a truncated Fourier representation of the local orography. Following \cref{eq:series}, this representation takes the form
\begin{equation}
h^{\left(j\right)}\left(x,y\right) = \sum_{m=-\mathcal{M}}^{\mathcal{M}}\sum_{n=-\mathcal{N}}^{\mathcal{N}} \hat{h}^{\left(j\right)}_{m,n}e^{i\left(k_m x + \ell_n y\right)},
\label{eq:tri_series}
\end{equation}
where the wavenumbers $k_m=2\pi m/L_x^{\left(j\right)}$ and $\ell_n=2\pi n/L_y^{\left(j\right)}$ are defined with respect to a local spectral window of size $L_x^{\left(j\right)}\times L_y^{\left(j\right)}$ associated with the triangle $T^{\left(j\right)}$. The formulation in \cref{sec:nufft_formulation_efficient_discrete_evaluation} applies to $T^{\left(j\right)}$ once periodicity is addressed. The Fourier series in \cref{eq:tri_series} implicitly assumes that $h^{\left(j\right)}\left(x,y\right)$ is periodic on the local window, which is rectangular. However, the triangle itself is not rectangular, and the DEM points are irregularly distributed within it.

To overcome this issue, local windowing and quadrature weights are introduced. Six window configurations are considered, obtained from two local frames and three support masks as described in \cref{app:local_windowing}. For each local window, two weighting schemes are considered as described in \cref{app:quadrature_weights}. Furthermore, the discrete spectral recovery fidelity is quantified and the possible sources of error are identified in \cref{app:discrete_spectral_fidelity}.

\subsection{Elastic Mode Selection}\label{sec:elastic_mode_count}
While the previous subsection proposed a framework for obtaining Fourier coefficients from scattered data on triangular grids without interpolation or fitting, this subsection develops a spectral window that elastically truncates less significant modes. The selection sufficiently represents the full spectrum while still respecting bounds set by a predefined spectral budget and launch-relevant power. Although developed to complement the NUFFT framework of \cref{sec:nufft}, Elastic Mode Selection (EMS) can be applied without bias to any spectrum, regardless of how the coefficients are obtained (e.g., a fitted spectrum from \citet{chew2024aconstrained}).

\subsubsection{Setup}
For a spectral mode $\left(m,n\right)$, its spectral energy can be defined as
\begin{equation}
\label{eq:energy}
E_{m,n}=\left|\hat h_{m,n}\right|^2.
\end{equation}
It may also be generalized to a mode pair $\pm\left(m,n\right)$ as
\begin{equation}
E_{m,n}=\left|\hat h_{m,n}\right|^2 + \left|\hat h_{-m,-n}\right|^2 ,
\end{equation}
and if launch physics weighs modes (e.g., by intrinsic frequency or vertical group velocity), a nonnegative weight $w_{m,n}$ can also be folded into the weighted energy as
\begin{equation}
\label{eq:weighted_e}
\widetilde E_{m,n}=w_{m,n}E_{m,n}.
\end{equation}
All selection rules below work with either $E$ or $\widetilde E$ for either signed modes or mode pairs. Let the set of energy spectra $E$ be indexed by $j=1,\dots,J^\star$, with its entries sorted in descending order as
\begin{equation}
  \label{eq:sorted_energy}
E_{\left(1\right)}\ge E_{\left(2\right)}\ge \cdots \ge E_{\left(J^\star - 1\right)} \ge E_{\left(J^\star\right)}.
\end{equation}
Set \cref{eq:sorted_energy} is then also the set of sorted importance. Hereafter, two complementary metrics are defined. The first is the participation ratio, which is based on equating moments, and the second is the gap ratio, which is based on neighborhood similarity. They both have complementary degeneracies and hence are combined into a single measure that best quantifies the shape of the spectrum and guides the selection of modes.

\subsubsection{Participation Ratio}\label{sec:participation_ratio}

Following the approach of \citet{recanatesi2022ascaledepend}, one can define $p_j=E_{\left(j\right)}/\left(\sum_{r}E_{\left(r\right)}\right)$ such that the participation ratio
\begin{equation}
\label{eq:n_eff}
N_{\mathrm{eff}}=\frac{1}{\sum_{j}p_j^2}
=\frac{\left(\sum_{j}E_{\left(j\right)}\right)^2}{\sum_{j}E_{\left(j\right)}^2}
\end{equation}
ranges from $1$ (one dominant line) up to $J^\star$ (all similar). If one mode dominates, then $p_1\approx 1$ and $N_{\mathrm{eff}}\approx 1$. If many modes share energy uniformly, $p_j\approx 1/J^\star$ and $N_{\mathrm{eff}}\approx J^\star$. This can be interpreted as a global notion of ``how many modes matter''.

\subsubsection{Gap Ratio}

On its own, \cref{sec:participation_ratio} is not enough to judiciously compress a given spectral distribution for optimum packing efficiency. This shall be discussed shortly. For now, one can additionally define the gap ratio $G_j={E_{\left(j\right)}}/{E_{\left(j+1\right)}}\ge 1$ (similar to eigen-gap-ratio in \citet{lee2021sourceenumer}) and the spectral clustering score (similar to the Gaussian similarity function in \citet{luxburg2007atutorialons})
\begin{equation}
\label{eq:s_delta}
S_\delta=\frac{1}{K_{\max}-1}\sum_{j=1}^{K_{\max}-1}\exp\left(-\frac{G_j-1}{\delta}\right),\qquad \delta>0.
\end{equation}
If neighbors are similar ($G_j\approx 1$), the exponential is near 1. A large gap ($G_j\gg 1$) pushes that term toward 0. Averaging only over the first $K_{\max}-1$ neighbors focuses this score where truncation actually occurs. If the top modes are tightly packed (flat top), then $G_j\approx 1$ and each term is near $1$, so $S_\delta$ is also near 1. A decisive gap ($G_j\gg1$) pushes that term toward $0$.

\subsubsection{Combined Measure}
\begin{figure}[!t]
\centering
\includegraphics[width=\textwidth]{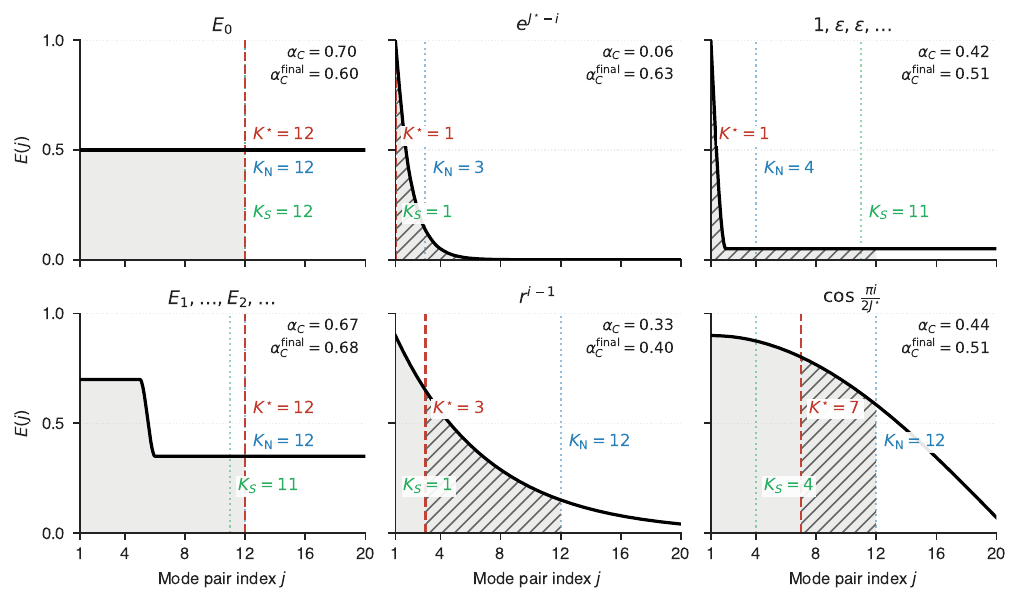}
\caption{Elastic mode selection applied to six spectral families including uniform, exponential, peak, step, geometric, and cosine. Here $J^\star=20$, $K_{\min}=1$, $K_{\max}=12$, $\alpha_{\min}=0$, $\alpha_{\max}=0.7$, $\delta=0.02$, and $w_1,w_2=0.5$. The vertical lines show the selected elastic count $K^\star$ (red-dashed), projected participation count $K_{\mathrm{N}}=N_{\mathrm{eff}}^{\left(\mathrm{clip}\right)}$ (blue-dotted), and projected similarity count $K_S=S_\delta K_{\max}$ (green-dotted). The target fraction $\alpha_C$ and final fraction $\alpha_C^{\mathrm{final}}$ constrained by \cref{eq:k_elastic} are also annotated. The areas under the curves are filled only up to the maximum elastic count $K_{\max}$, and the modes rejected by EMS are hashed.}
\label{fig:ems_spectra}
\end{figure}

As explained in \cref{app:ems_shortcomings}, both metrics struggle to recognize certain spectral shapes, and their shortcomings are somewhat complementary. Combining them solves this issue. First, $N_{\mathrm{eff}}$ is clipped by the hard budget, yielding
\begin{equation}
N_{\mathrm{eff}}^{\left(\mathrm{clip}\right)}=\min\left\{N_{\mathrm{eff}},K_{\max}\right\}.
\end{equation}
This is then linearly combined with the spectral clustering score using the weights $w_1,w_2 \in \left[0, 1\right]$ with $w_1+w_2=1$, so that the result is the normalized combined measure
\begin{equation}
\label{eq:c_score}
\mathcal{C}=w_1\frac{N_{\mathrm{eff}}^{\left(\mathrm{clip}\right)}}{K_{\max}} + w_2S_\delta \quad \in \left[0,1\right],
\end{equation}
where $\mathcal{C}\to1$ implies a ``flat head'' spectrum (broadband), while $\mathcal{C}\to0$ implies a ``sharp head'' spectrum (narrowband). To determine the selected-mode count $K^\star$, $\mathcal{C}$ is mapped to a target retained-power fraction $\alpha\in\left(0,1\right)$ as
\begin{equation}
\label{eq:alpha_c}
\alpha_C=\alpha_{\min}+\left(\alpha_{\max}-\alpha_{\min}\right)\mathcal{C}
\end{equation}
with $0<\alpha_{\min}<\alpha_{\max}<1$ \footnote{This algorithm assumes $K_{\max}<J^{\star}$. Although this condition is realistic in practice, computational robustness is improved by using $J_{\max}=\min\left\{J^{\star}, K_{\max}\right\}$ as the upper bound for all measures and sums. This technical detail does not affect the main idea.} and, ultimately, a retained fraction
\begin{equation}
\label{eq:f_cum}
\vartheta\left(K\right)=\frac{\sum_{j=1}^{K}E_{\left(j\right)}}{\sum_{r=1}^{J^\star}E_{\left(r\right)}}
\end{equation}
such that the elastic count can be obtained as
\begin{equation}
\label{eq:k_elastic}
K^\star=\min\left\{K_{\max},\max\left\{K_{\min},\min\left\{K\,:\,\vartheta\left(K\right)\ge \alpha_C\right\}\right\}\right\}.
\end{equation}
This choice thus preserves the leading spectral shape while enforcing the spectral budget $K_{\min}\le K^\star\le K_{\max}$ and, whenever permitted by that budget, satisfying the retained-power target $\vartheta\left(K^\star\right)\ge\alpha_C$. The full elastic-mode-selection algorithm is illustrated in \cref{fig:ems_spectra} for different spectral families, including cases where the participation ratio or gap ratio alone would suffer degeneracies\footnote{
The exact configurations are as follows. Uniform has $E=0.5$ for all modes. Exponential has $E=e^{1-j}$ for all modes. Peak has $E=1$ for the first mode and $\varepsilon=0.05$ for the rest. Step has $E=0.7$ for the first five modes and $0.35$ for the rest. Geometric has $E=0.9\left(0.85\right)^{j-1}$ for all modes. Cosine has $E=0.9\cos\left[\pi\left(j-1\right)/\left(2J^\star\right)\right]$ for all modes.}.

\subsection{Comparison to Constrained Spectral Approximation}\label{sec:csa_comparison}
As the current most applicable approach for irregular DEMs on unstructured grids, the Constrained Spectral Approximation method (CSA) of \citet{chew2024aconstrained} is used as a reference for this study. For consistency with the published method description, the comparison uses the LSFF/FA--SA formulation and regularization values reported by \citet{chew2024aconstrained}, with both CSA and ENUFFT spectra expressed in the same signed-complex Fourier convention. Practical CSA implementations may additionally use tapering and scaled regularization, but these choices are not required by the published formulation and are hence not included in the default comparison here. A detailed comparison is provided in \cref{app:csam_comparison}. The key relationship can be expressed as
\begin{equation}
  \hat{ h}^{\mathrm{CSA}}
  =  M^{-1}\hat{ h}^{\mathrm{ENUFFT}},
  \qquad
   M = \frac{1}{Q}\left( F^{\ast} F + \lambda I_J\right),
\end{equation}
where $F$ is the Fourier transform matrix (see \cref{app:csam_comparison}). Therefore, CSA coefficients can be interpreted as a filtered version of the ENUFFT coefficients with the filter given by the inverse normal matrix $M$. This implies that on top of the differences arising from approximation through fitting versus quadrature, the CSA coefficients also have all the truncation and regularization effects embedded into its coefficients. Meanwhile, the ENUFFT coefficients represent a more direct approximation of the Fourier transform, and all truncation-related effects are handled separately by the elastic mode selection defined in \cref{sec:elastic_mode_count}. This is the fundamental difference between the two approaches. Unlike CSA, ENUFFT can dynamically change the choice or number of truncated modes without recomputing the coefficients. Furthermore, as shown in \cref{app:csam_enufft_complexity}, for $N$ Fourier modes in each direction and $Q$ DEM points, CSA's compute cost approximately scales as
\begin{equation}
  \mathrm{CSA\ cost} =\mathcal{O}\left(QN^4 + N^6\right)
\end{equation}
while that of ENUFFT scales as
\begin{equation}
  \mathrm{ENUFFT\ cost} = \mathcal{O}\left(Q + N^2 \log N\right).
\end{equation}
Both scale linearly with $Q$, but ENUFFT has more favorable scaling with $\sim N^2\log N$ compared with CSA's $\sim N^6$.
\section{Application}
This section demonstrates the performance of ENUFFT in both synthetic and real-world scenarios. It is broadly divided into two parts - static or flow-independent, and dynamic or flow-dependent. The former focuses on ENUFFT's ability to represent terrain spectra on unstructured grids and is further divided into synthetic monochromatic and real Alpine terrain tests but under static conditions. The latter meanwhile further demonstrates the ability of EMS and by extension ENUFFT, to not just represent terrain spectra but also to judiciously select the most important modes for launch under real-time flow conditions. To put everything into perspective, comparisons with the CSA method are also provided (where applicable).

\subsection{Monochromatic Test}\label{exp:monochromatic_test}
In this test, ENUFFT is applied to a setup whose spectrum is known. Because fitting-based approaches such as CSA are motivated by the difficulty of reconstructing truly local signals from few DEM samples on a triangle via quadrature, a locally varying monochromatic signal serves as the harshest test case. Let the local orography be
\begin{equation}
    h^{\left(j\right)}\left(x',y'\right) = A\cos\left(\frac{2\pi m^{\left(j\right)}_s}{L_x^{\left(j\right)}}x' + \frac{2\pi n^{\left(j\right)}_s}{L_y^{\left(j\right)}}y' + \phi^{\left(j\right)}_s\right),
    \label{eq:mono_local_p2}
\end{equation}
where $j$ is the triangle index, $L_x,L_y$ the edge lengths of the respective window domain (see \cref{app:local_windowing}), and $\phi^{\left(j\right)}_s$ a deterministic random phase assigned to that local signal. Because a window around a triangle encroaches on the monochromatic signal of its neighbors, this setup pushes ENUFFT to its limit. Implementation details are provided in \citet{banerjee2026enufftsoftware,banerjee2026elasticnufftgithub} but in summary, the domain consists of a square $D=\left[0,L\right]\times\left[0,L\right]$ with a triangle $T$ and a circle $C$ inscribed in it. The triangle can move vertically (offset $\Delta y$), rotate around its center (orientation $\theta_T$), modulate its aspect ratio (uniformity $u$), and scale its size (expansion ratio $r_{\mathrm{exp}}$). Throughout the domain, $Q$ DEM points are irregularly distributed. Furthermore, $D$ is divided into a maximum of four regions, namely $T$ itself and up to the three sectors into which $T$ divides the rest of $D$ if the lines connecting its vertices to its center are extended to intersect $D$.

By modulating the triangle's position, orientation, aspect ratio, and size, the test can mimic any triangular-grid-cell configuration of an unstructured grid, where the bounding square is the local window. Similarly, the square, the embedded circle, or the triangle itself can serve as additional window masks for such a configuration, effectively enforcing the support choices described in \cref{app:local_windowing}. This setup enables a comprehensive and affordable parameter sweep across all possible unstructured-grid-cell configurations. The parameter vector is defined by
\begin{equation}
    \mathbf{p} = \left(\Delta y, \theta_T, u, r_{\mathrm{exp}}, Q, \sigma, K_{\max}, w_q\right),
\end{equation}
where $\sigma$ is the ENUFFT grid oversampling factor, $K_{\max}$ the maximum number of mode pairs allowed, and $w_q$ the weight for each DEM point, with the sweep configurations listed in \cref{tab:mono_sweep_config}. The full sweep results can be found in \cref{tab:mono_direction_sweep,tab:mono_amplitude_sweep,tab:mono_budget_sweep} and the raw dataset in \citet{banerjee2026enufftdataset}.
\begin{table}
\caption{Parameter-sweep configurations for the monochromatic test.}
\label{tab:mono_sweep_config}
\centering
\begin{tabular}{p{0.42\textwidth}p{0.48\textwidth}}
\hline
Parameter & Values\\
\hline
Vertical triangle offset $\Delta y/L$ & $-0.211$, $-0.077$, $0$ \\
Triangle orientation $\theta_T$ & $0\,^\circ$, $60\,^\circ$, $120\,^\circ$, $180\,^\circ$, $240\,^\circ$, $300\,^\circ$ \\
Triangle uniformity $u$ & $0$, $0.5$, $1$ \\
Expansion ratio $r_{\mathrm{exp}}$ & $1$, $1.5$, $2$ \\
DEM point count $Q$ & $500$, $1000$, $2000$ \\
ENUFFT oversampling $\sigma$ & $1.25$, $1.5$, $2$ \\
Maximum retained mode pairs $K_{\max}$ & $4$, $6$, $8$, $10$ \\
DEM point weights $w_q$ & uniform, Voronoi area \\
Window mask & triangle, circle, square \\
\hline
\end{tabular}
\end{table}
\begin{figure}[!t]
\centering\includegraphics[width=\textwidth]{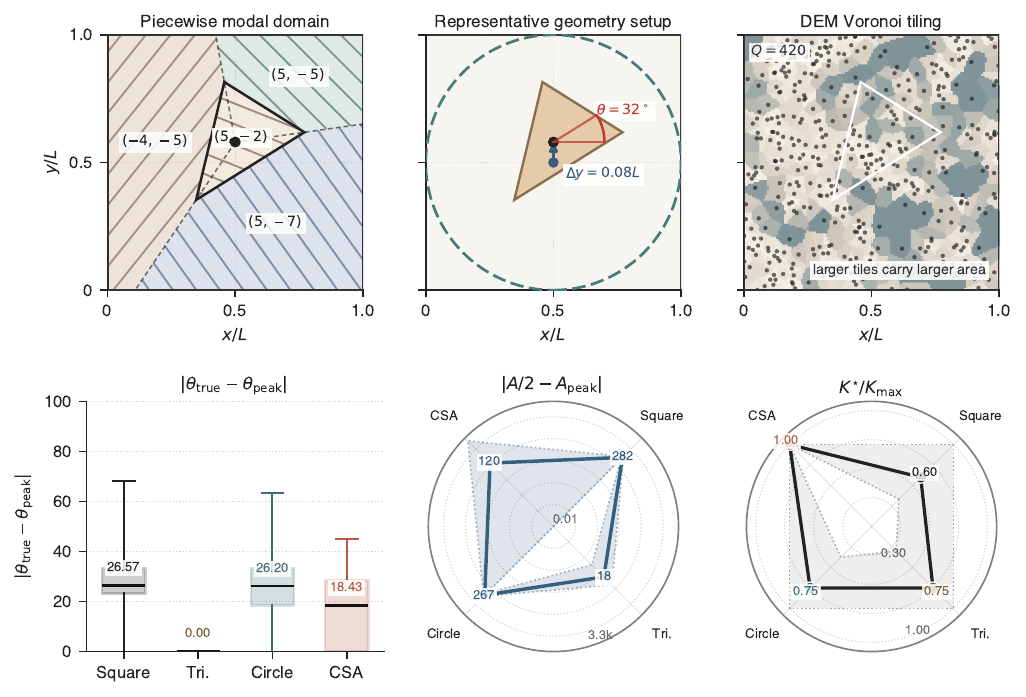}
\caption{Monochromatic test of ENUFFT and CSA \cref{exp:monochromatic_test} with $A=1000$ m. The top row shows the test setup and geometry, while the bottom row shows pooled results (peak direction recovery $\Delta\theta$, peak amplitude recovery $\Delta A$, and effective spectral budget compaction $K^\star/K_{\max}$). In all cases, solid lines indicate median values. Box-plot filled areas show interquartile ranges (25th, 75th percentiles), while whiskers show 10th, 90th percentiles. Radar-plot filled spreads also show 10th, 90th percentiles. ENUFFT uses the same configuration as \cref{fig:ems_spectra}, with $\gamma,\beta$ given by \cref{eq:kaisal_bessel_coeffs}, while CSA follows \citet{chew2024aconstrained}, with $\lambda_{\mathrm{FA}}=10^{-1}$ and $\lambda_{\mathrm{SA}}=10^{-6}$.}
\label{fig:mono_test}
\end{figure}
The test setup and resulting sweep are illustrated in \cref{fig:mono_test}. Overall, ENUFFT produces peak signal recovery comparable to CSA in both magnitude and direction while using less than the full spectral budget. More specifically, ENUFFT with the triangular window recovers $100\,\%$ of the peak signal mode, compared with CSA's median $18.43\,^\circ$ error, and also performs better than CSA in peak signal amplitude recovery (median error of $18\,\mathrm{m}$ versus $120\,\mathrm{m}$ for CSA, for the physical amplitude $A/2=500\,\mathrm{m}$). Regardless of the parameter-sweep configuration, ENUFFT also consistently uses fewer spectral modes (median \allowbreak $K^\star/K_{\max}\leq0.75$). The $10$th--$90$th percentiles further favor ENUFFT as the CSA direction recovery error rises to $\sim 55\,^\circ$, and the amplitude recovery error to $\sim 3300\,\mathrm{m}$. ENUFFT with the triangular window, meanwhile, maintains $100\,\%$ direction recovery and has a maximum amplitude error of $\sim 60\,\mathrm{m}$.

\subsection{Alpine Test}\label{exp:alps_real_terrain_test}
\begin{figure}[!t]
\centering
\includegraphics[width=\textwidth]{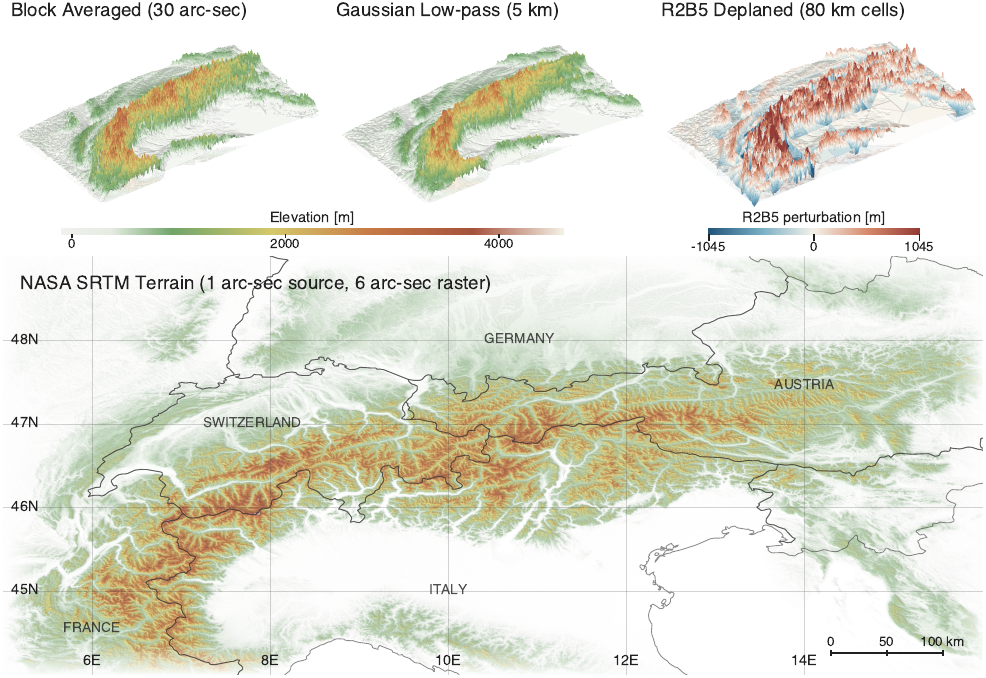}
\caption{Alpine SRTM DEM of \cref{exp:alps_real_terrain_test} through four stages of preprocessing, namely raw SRTM mosaic, coarse-grained with elevation clipping, Gaussian smoothed, and deplaned on an R2B5 mesh. The source topography is the NASA SRTM GL1 1 arc-second dataset \citep{nasajpl2013srtmgl1}.}
\label{fig:alps_preprocessing_viz}
\end{figure}

Having investigated the recovery of an analytic monochromatic orography, the ENUFFT pipeline is next applied to a real Alpine DEM from the NASA Shuttle Radar Topography Mission (SRTM) Global 1 arc-second product \citep{nasajpl2013srtmgl1}, whose mission design and topographic data characteristics are described by \citet{farr2007shuttleradart}. The domain spans latitudes from $44\,^\circ\mathrm{N}$ to $49\,^\circ\mathrm{N}$ and longitudes from $5\,^\circ\mathrm{E}$ to $16\,^\circ\mathrm{E}$, encompassing the full Alpine arc. First, the DEM is preprocessed in four broad steps - base-elevation clipping, block averaging, Gaussian spectral smoothing, and deplaning, as shown in \cref{fig:alps_preprocessing_viz} and expanded in \cref{app:alps_preprocessing}. \citet{banerjee2026enufftsoftware,banerjee2026elasticnufftgithub} documents the test implementation, including the preprocessing, in detail.
The mesh-dependent deplaning is performed after the DEM has been sampled on two deterministic ICON-like meshes (R2B4 and R2B5)\footnote{The coarser R2B4-scale proxy mesh uses target cell size $\Delta=160\,\mathrm{km}$, square mode limit $\mathcal{M}=\mathcal{N}=16$, and 48 triangles. The finer R2B5-scale proxy mesh uses $\Delta=80\,\mathrm{km}$, $\mathcal{M}=\mathcal{N}=32$, and 154 triangles.}. On each mesh, the same sweep from \cref{tab:mono_sweep_config} is performed\footnote{There are some technical differences between the sweep defined by \cref{tab:mono_sweep_config} and that in \cref{exp:alps_real_terrain_test}. The latter excludes the parametric sweep in $\theta_T$, $u$, $Q$, $K_{\max}$, and $\Delta y/L,$.}. The full sweep results for both meshes can be found in \cref{tab:alps_r2b4_rmse,tab:alps_r2b4_budget,tab:alps_r2b4_variance,tab:alps_r2b5_rmse,tab:alps_r2b5_budget,tab:alps_r2b5_variance} and the raw dataset in \citet{banerjee2026enufftdataset}.
\begin{figure}[!t]
\centering
\includegraphics[width=\textwidth]{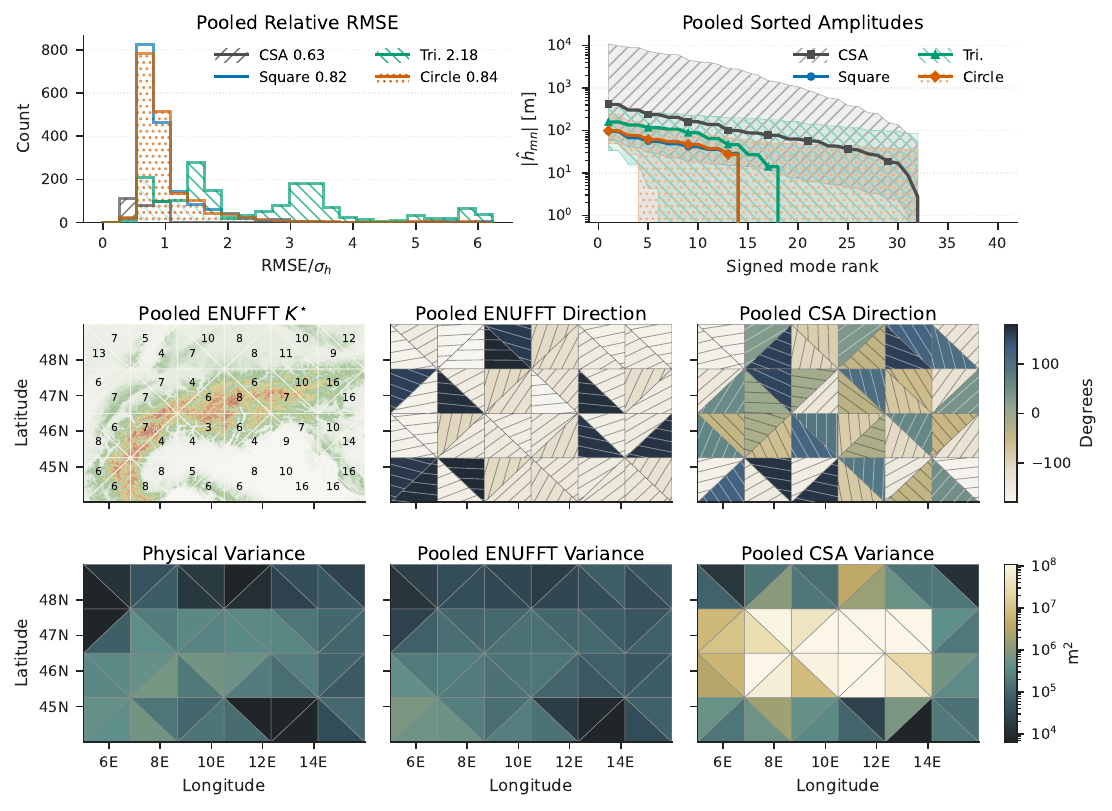}
\caption{ENUFFT and CSA applied to the Alpine SRTM DEM from \cref{exp:alps_real_terrain_test} on an irregular triangular mesh (R2B4 $\Delta\approx160$ km). All plots show pooled median results. The top row shows relative reconstructed RMSEs and sorted spectral amplitudes. The spread shows 10th, 90th percentiles. The middle row shows ENUFFT $K^\star$ and ENUFFT, CSA dominant-mode direction distributions. The bottom row shows physical DEM variance and spectral variance distributions for both methods. ENUFFT uses the same configuration as in \cref{fig:ems_spectra}, with $\gamma,\beta$ given by \cref{eq:kaisal_bessel_coeffs}, while CSA follows \citet{chew2024aconstrained}, with $\lambda_{\mathrm{FA}}=10^{-1}$ and $\lambda_{\mathrm{SA}}=10^{-6}$.}
\label{fig:alps_sweep_r2b4}
\end{figure}
\begin{figure}[!t]
\centering
\includegraphics[width=\textwidth]{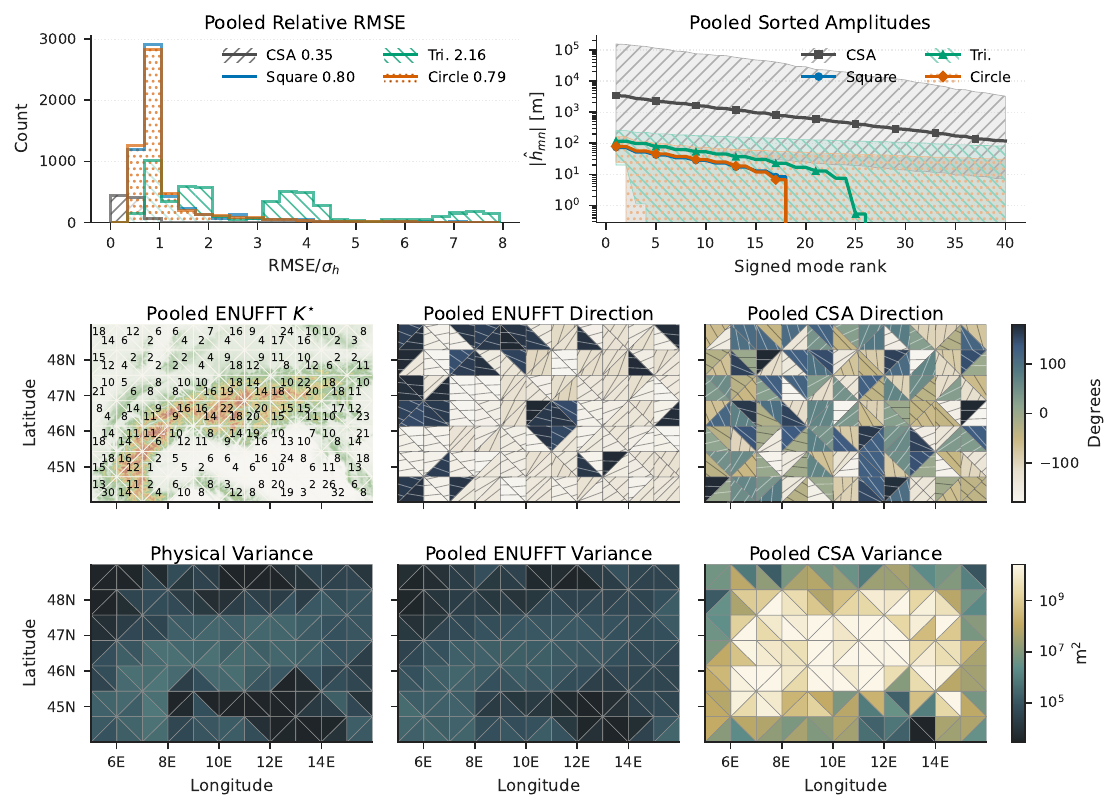}
\caption{Same as \cref{fig:alps_sweep_r2b4} but on the finer R2B5 mesh ($\Delta\approx80$ km).}
\label{fig:alps_sweep_r2b5}
\end{figure}

\cref{fig:alps_sweep_r2b4,fig:alps_sweep_r2b5} show the sweep results. Overall, ENUFFT once again produces results comparable to those of CSA while drastically compacting the spectrum and preventing spurious energy from being injected into it. More specifically, in terms of the median relative reconstructed physical RMSE per triangle, defined by
\begin{equation}
  \varepsilon_{\mathrm{rel}}
  =
  \frac{
  \left\{
  \left(Q^{\left(j\right)}\right)^{-1}\sum_{q\in T^{\left(j\right)}}
  \left[h'_q - h_{\rm rec}\left(x_q,y_q\right)\right]^2
  \right\}^{1/2}
  }{
  \left[
  \left(Q^{\left(j\right)}\right)^{-1}\sum_{q\in T^{\left(j\right)}}
  \left(h'_q\right)^2
  \right]^{1/2}
  },
  \label{eq:alps_relative_rmse}
\end{equation}
where $h'_q$ is defined in \cref{eq:alps_deplaning}, CSA performs better. This is expected, since the CSA objective \cref{eq:csam_objective} specifically targets fitting against the physical coefficients. Meanwhile, ENUFFT still produces a median $\varepsilon_{\mathrm{rel}}$ only $\sim 0.3$--$0.5$ standard deviations worse than CSA. At the same time, ENUFFT uses significantly fewer modes, with median spectral compaction ($K^\star$) down to just $\sim30$--$50\,\%$ of the CSA count (i.e., the spectral budget $K_{\max}$). Furthermore, although reconstructed RMSE is a valid comparison metric, it remains a physical-space measure. While both ENUFFT and CSA provide spectra on each mesh triangle, there is no straightforward reference spectrum for verification. Without interpolation in physical or spectral space, the most direct test is then Parseval's theorem, which states that the total energy of a signal is the same whether computed in the physical domain or the spectral domain, i.e.,
\begin{equation}
\sigma^2_{\mathrm{phys}} = \frac{1}{\left|D\right|}\sum_{q} w_q \left(h'_q\right)^2 = \sum_{m,n\neq 0,0} \left|\hat{h}_{m,n}\right|^2 = \sigma^2_{\mathrm{spec}}.
\label{eq:spectral_variance}
\end{equation}
\citet{Chen_2024} demonstrated this relationship in detail in a discrete energy-conservation setting. As shown in the last-row plots of \cref{fig:alps_sweep_r2b4,fig:alps_sweep_r2b5}, ENUFFT comes close to satisfying the Parseval condition (median deviation R2B4 $\sim25\,\%$, R2B5 $\sim14\,\%$), whereas CSA greatly overshoots it (median deviation R2B4 $>500\,\%$, R2B5 $>122{,}000\,\%$) by injecting more spurious energy into the reconstructed spectra than the physical orography supports, especially in the central Alps. This behavior is a typical indication of overfitting. Although the regularizers in CSA could be tuned to improve the results, the comparison also exposes the higher sensitivity of fitting-based methods relative to ENUFFT, which has not been tuned either. Finally, the triangular window for ENUFFT performs worst, in direct contrast to the monochromatic test in \cref{exp:monochromatic_test}. This result can be interpreted as an intrinsic limitation of the triangular window. Its discontinuous boundary and lower DEM count, compared with other windows, can be overcome only for highly distinctive, singular, and localized source spectra (like the monochromatic test of \cref{exp:monochromatic_test}).

\subsection{Mountain-Wave Test}\label{exp:mountain_wave}
Finally, as mentioned in \cref{sec:elastic_mode_count}, the elastic-mode-selection algorithm (EMS) can be applied not only statically but also dynamically. To illustrate this, EMS was implemented in the idealized atmospheric-flow solver PinCFlow.jl \citep{jochum2026pincflow} to supplement the orographic source in its implementation of the next-generation 3D transient gravity-wave parameterization MS-GWaM \citep{Muraschko2015, Boeloeni2016, Boeloeni2021, Kim2021, Kim2024, Voelker2024, Jochum2025}. Because MS-GWaM models gravity waves using computationally expensive propagating ray volumes, it is well suited for demonstrating the potential of EMS. Here, EMS is used in an idealized one-hour simulation of gravity waves generated above an isolated mountain range.

\citet{banerjee2026enufftsoftware,banerjee2026elasticnufftgithub} describes the test setup in detail, including the modified \citet{atmosphericdynamicsguf2026pincflowgithub} implementation used here but, in summary, the configuration features a domain $D=\left[-L_x/2,L_x/2\right]\times\left[-L_y/2,L_y/2\right]\times\left[0,L_z\right]$ with orography defined by
\begin{align}\label{eq:mountain_orog_1}
h \left(x, y\right) = h_{\symup{b}} \left(x, y\right) + \sum\limits_{\mu = 1}^{N_{\mu}}\hat{h}_{\symup{w},\mu} \left(x, y\right) \cos \left(k_\mu x + \ell y\right),
\end{align}
where the resolved background $h_{\symup{b}}$, wavenumbers $\left(k_{\mu}, \ell\right)$, and envelope $\hat{h}_{\symup{w},\mu}$ are given by
\begin{align}\label{eq:mountain_orog_2}
h_{\symup{b}} \left(x, y\right) & = N_\mu r_h h_{\symup{w}} \left(x, y\right), \quad k_\mu = \frac{2 \pi}{\lambda_0} \left(\mu - 1\right), \ell = \frac{2 \pi}{\lambda_0}\\
\hat{h}_{\symup{w},\mu} \left(x, y\right) & = \begin{cases}
    \frac{h_0}{2 N_\mu \left(r_h + 1\right)} \left[1 + \cos \left(\frac{2 \pi}{r_\lambda \lambda_0} \sqrt{x^2 + y^2}\right)\right], & \quad x^2 + y^2 \leq \frac{r_\lambda^2 \lambda_0^2}{4},\\
    0, & \quad x^2 + y^2 > \frac{r_\lambda^2 \lambda_0^2}{4}.
\end{cases}
\end{align}
Here, $h_0$ denotes maximum height, $\lambda_0$ the largest unresolved wavelength, $r_h$ the resolved-to-summed unresolved amplitude ratio, $r_\lambda$ the resolved-to-unresolved wavelength ratio, and $N_\mu$ the number of modes. The wind is initialized with $\mathbf{v}_0=\left(u_0,0,0\right)$. As explained in \cref{app:mountain_wave}, the wave-action density of an orographic gravity wave depends not only on orography but also on the flow as\footnote{Note that \cref{eq:mw_action_2} leads to a signed wave-action density and $\symscr{A}_\mu,\hat{\omega}_\mu$ are always chosen from the same signed branch to ensure positive wave energy $E_\mu=\symscr{A}_\mu\hat{\omega}_\mu$.},
\begin{equation}
\symscr{A}_\mu
=-
\frac{\bar{\rho}}{2}
\mathbf{k}_{\mu}\cdot\mathbf{u}_b
\frac{\mathbf{k}_\mu^2+m_\mu^2}{\mathbf{k}_\mu^2}
\left|h_{\symup{w},\mu}\right|^2
\label{eq:mw_action_2}
\end{equation}
In \cref{eq:mw_action_2}, $h_{\symup{w},\mu}$ is the complex modal amplitude used in \cref{app:mountain_wave}. For the real cosine terrain in \cref{eq:mountain_orog_1}, $\left|h_{\symup{w},\mu}\right|=\left|\hat{h}_{\symup{w},\mu}\left(x,y\right)\right|$. This implies $\symscr{A}_1 = 0$ at the initial time because $\left(k_1,\ell\right)=\left(0,2\pi/\lambda_0\right)$. If EMS prevents ray-volume launches for modes with low $\left|\symscr{A}_\mu\right|$, it can save computational resources with little loss of potential mean-flow impact.
\begin{figure}[!t]
\centering
\includegraphics[width=\textwidth]{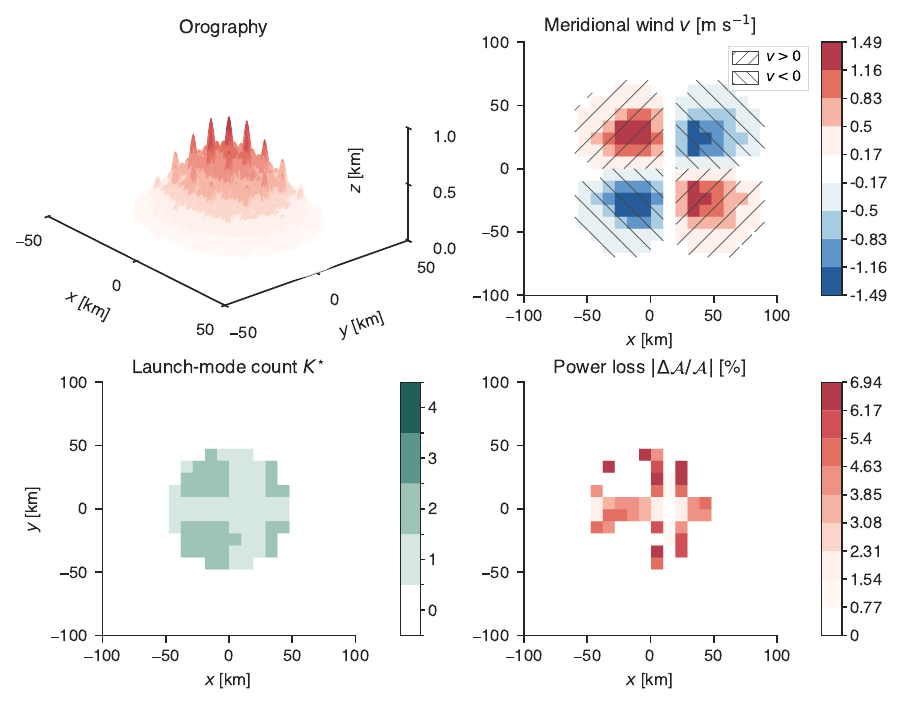}
\caption{Results of a one-hour PinCFlow.jl simulation from \cref{sec:elastic_mode_count}, with EMS applied to the orographic wave-action-density spectrum. EMS specifications are as in \cref{fig:ems_spectra} except for $\alpha_{\min}=0.9$ and $\alpha_{\max}=1.0$. The simulation is conducted in the domain $D=\left[-100,100\right]\times\left[-100,100\right]\times\left[0,10\right]\,\mathrm{km}^3$ with grid dimensions $\left(n_x,n_y,n_z\right)=\left(20,20,10\right)$ and spectral budget $K_{\max}=N_{\mu}=4$. The atmosphere is pseudo-incompressible, with a constant-temperature-gradient background and initial zonal wind $u_0=10\,\mathrm{m\,s}^{-1}$. Orography is given by \cref{eq:mountain_orog_2} with $r_\lambda=10$, $r_h=2$, $h_0=1\,\mathrm{km}$, and $\lambda_0=10\,\mathrm{km}$. Plotted are spatial distributions of orography, meridional surface wind (with directional hatching), launch-mode count $K^\star$ under EMS, and relative launch-power loss $\Delta\symscr{A}/\symscr{A}=1-\sum_{k=1}^{K^\star}\symscr{A}_k/\sum_{k=1}^{K_{\max}}\symscr{A}_k$.}
\label{fig:pincflow_ems}
\end{figure}

\cref{fig:pincflow_ems} shows the results of a one-hour simulation, the raw dataset of which can be found in \citet{banerjee2026enufftdataset}. The top-right panel shows a meridional surface wind induced as the mountain diverts the flow, slightly flattening wave-action-density distributions in affected grid cells. However, at $y = 0\,\mathrm{km}$ and $x\in[10,20]\,\mathrm{km}$, the meridional wind is comparatively weak, so mode selection in this region is not substantially affected. Most noticeably, the launch-mode count $K^\star$ does not exceed 2, whereas the launch-power loss remains below $7\,\%$. This corresponds to significant compaction (median compaction $\sim75\,\%$ for all cells with $K^\star>0$) with a relatively small penalty. Not only are the results consistent with the expected behavior for a configuration with uniform $h_{\symup{w}}$ across modes combined with a relatively large factor between the largest and smallest unresolved wavenumbers, they illustrate the effectiveness and necessity of dynamic mode selection.

\section{Summary and Conclusion}
This study proposed Elastic Non-Uniform Fast Fourier Transform (ENUFFT) for obtaining spectra from irregularly sampled data on non-uniform triangular grids. The approach extended global type-1 NUFFT to local triangular meshes and introduced Elastic Mode Selection (EMS) to dynamically compress the resulting spectra while minimizing loss of power or fidelity. The pipeline was tested with both synthetic and real-world data \cref{exp:monochromatic_test,exp:alps_real_terrain_test} and against the current best alternative approach in Constrained Spectral Approximation (CSA) of \citet{chew2024aconstrained}. ENUFFT performed comparably to or better than CSA across the considered metrics. For amplitude-related errors (physical and spectral), ENUFFT and CSA produced broadly comparable results, with ENUFFT performing slightly better in the synthetic-data test but slightly worse in the real-world-data test. More importantly, ENUFFT achieved comparable results while compacting the spectra by $60\,\%$ on average and also maintaining significantly better energy conservation. As shown in the real-world-data test \cref{exp:alps_real_terrain_test}, CSA can produce median spurious spectral variance (energy) 3 orders of magnitude higher than what the physical orography can support, whereas ENUFFT maintains a difference below 1 order of magnitude. EMS was also applied to a dynamic spectrum in an idealized simulation with a 3D transient mountain-wave parameterization. This highlighted its ability to compress any given spectrum intelligently and dynamically. In the idealized simulation, EMS compressed the gravity-wave launch spectrum in a flow-dependent manner by $\ge75\,\%$ on average while maintaining $\leq7\,\%$ power loss. Furthermore, the tests showed that ENUFFT performs comparably or better than CSA in signal direction recovery as well. Overall, the study demonstrated ENUFFT to be a promising new way to obtain spectra from irregularly sampled orography on non-uniform triangular meshes.

\section{Outlook}\label{sec:outlook}
For potential future outlook, the two main pieces of the framework can be developed separately. EMS only needs a finite, ranked set of modal weights, so it is not tied to ENUFFT or even to orographic gravity waves. It could be applied to spectra obtained from CSA, from another source model, or from any ray-launch problem in which a local and possibly time-dependent spectrum must be compressed while controlling the loss of a chosen physical weight. In this sense, EMS is best viewed as a spectral-budget allocator rather than only as a post-processing step for NUFFT-derived terrain spectra.

The NUFFT part also has uses beyond topography. Turbulence and geophysical-flow studies often need local spectra to diagnose scale content, but Fourier analysis on triangular or otherwise unstructured meshes is sensitive to interpolation choices \citep{juricke2023scaleanalysis}. A closely related use case is Lagrangian or meshless simulation data, where spectra must be estimated from scattered particle samples \citep{shi2013interpolation}. A local NUFFT formulation can instead recover coefficients directly from irregular samples, after which the user can apply arbitrary spectral operations such as hard-cutoffs, band-pass filters, directional masks, or targeted splicing. The same construction should also extend beyond triangles to other polygonal cells, including hexagonal grids, as long as a local rectangular Fourier domain, support mask, and appropriate quadrature weights can be defined.

\section*{Open Research}
The dataset and analysis code for all experiments in this study are publicly available. The NASA Shuttle Radar Topography Mission Global 1 arc-second data used for the Alpine experiments is available from the NASA Land Processes Distributed Active Archive Center \citep{nasajpl2013srtmgl1}. Codes used for analyses and plotting are archived at \citet{banerjee2026enufftsoftware}. Datasets supporting the reported experiments and figures are archived at \citet{banerjee2026enufftdataset}. Detailed implementation, including a modified version of the \citet{atmosphericdynamicsguf2026pincflowgithub} setup used for the mountain-wave EMS test, is available at \citet{banerjee2026elasticnufftgithub}. Licenses and access conditions are given in the cited repository records.


\section*{Conflict of Interest}
The authors declare there are no conflicts of interest for this manuscript.

\acknowledgments
 T.B., F.J., and U.A. thank the Deutsches Klimarechenzentrum (DKRZ) for compute resources granted by Scientific Steering Committee (WLA) under project 1097 ``Multiscale Dynamics of Atmospheric Gravity Waves''. T.B., and U.A. thank the German Research Foundation (DFG) for partial support through CRC 301 ``TPChange'' (Project No. 428312742 and Projects B06 ``Impact of small-scale dynamics on UTLS transport and mixing'', and Z03 ``Joint model development and modelling synthesis''). T.B., and U.A. also thank DFG for support through the CRC 181 ``Energy transfers in Atmosphere and Ocean'' (Project No. 274762653 and Projects W01 ``Gravity-wave parameterization for the atmosphere'', and S02 ``Improved Parameterizations and Numerics in Climate Models''). F.J., and U.A. further acknowledge the support received through Schmidt Sciences. T.B also thanks Daniel Kunkel, Johannes Gutenberg University Mainz for making the SRTM DEM data available.

\bibliographystyle{plainnat}
\bibliography{References}

\clearpage
\appendix
\setcounter{section}{0}
\renewcommand{\thesection}{A\arabic{section}}
\renewcommand{\thesubsection}{\thesection.\arabic{subsection}}
\renewcommand{\thesubsubsection}{\thesubsection.\arabic{subsubsection}}
\crefalias{section}{appendix}
\crefalias{subsection}{appendix}
\crefalias{subsubsection}{appendix}
\crefalias{paragraph}{appendix}
\crefalias{subparagraph}{appendix}
\titleformat{\section}
  {\normalfont\large\bfseries}{Appendix \thesection}{0.75em}{}

\numberwithin{equation}{section}
\makeatletter
\@addtoreset{figure}{section}
\@addtoreset{table}{section}
\makeatother
\renewcommand{\theequation}{a\arabic{section}.\arabic{equation}}
\renewcommand{\thefigure}{a\arabic{section}.\arabic{figure}}
\renewcommand{\thetable}{a\arabic{section}.\arabic{table}}


\section{Local Windowing}\label{app:local_windowing}
This section describes the local windows used in the ENUFFT calculations. In the present implementation, each local Fourier basis is evaluated on a square spectral domain. The active support inside this square can then be square, triangular, or circular. Thus, the square domain fixes the Fourier periods, while the support mask fixes which DEM samples and which quadrature area enter the coefficient sum. For a triangle $T^{\left(j\right)}$ with vertices $\mathbf{v}_1^{\left(j\right)}$, $\mathbf{v}_2^{\left(j\right)}$, and $\mathbf{v}_3^{\left(j\right)}$, the centroid is
\begin{equation}
\bar{\mathbf{x}}^{\left(j\right)}
=
\frac{1}{3}\sum_{i=1}^{3}\mathbf{v}_i^{\left(j\right)}.
\end{equation}
Two orientations of the square domain are used. In the centroid-aligned case, the original local axes are retained,
\begin{equation}
\mathbf{R}^{\left(j,\mathrm{centroid}\right)} = I.
\end{equation}
In the edge-aligned case, the coordinate system is rotated so that the longest triangle edge becomes parallel to the local $x$-axis.\label{app:edge_aligned} The three edge vectors are
\begin{equation}
\mathbf{e}_1^{\left(j\right)} = \mathbf{v}_2^{\left(j\right)} - \mathbf{v}_1^{\left(j\right)}, \qquad
\mathbf{e}_2^{\left(j\right)} = \mathbf{v}_3^{\left(j\right)} - \mathbf{v}_2^{\left(j\right)}, \qquad
\mathbf{e}_3^{\left(j\right)} = \mathbf{v}_1^{\left(j\right)} - \mathbf{v}_3^{\left(j\right)}.
\end{equation}
The longest of these is denoted by $\mathbf{e}_{\max}^{\left(j\right)}=\left(e_x^{\left(j\right)},e_y^{\left(j\right)}\right)$. The corresponding angle of this oriented edge relative to the positive $x$-axis and its rotation matrix are then,
\begin{equation}
\cos\theta^{\left(j\right)}
=
\frac{e_x^{\left(j\right)}}{\left\|\mathbf{e}_{\max}^{\left(j\right)}\right\|},
\qquad
\sin\theta^{\left(j\right)}
=
\frac{e_y^{\left(j\right)}}{\left\|\mathbf{e}_{\max}^{\left(j\right)}\right\|}.
\qquad
\mathbf{R}^{\left(j, \mathrm{edge}\right)}
=
\begin{pmatrix}
\cos\theta^{\left(j\right)} & \sin\theta^{\left(j\right)} \\
-\sin\theta^{\left(j\right)} & \cos\theta^{\left(j\right)}
\end{pmatrix}.
\label{eq:rotation_matrix}
\end{equation}
For either orientation $a\in\{\mathrm{centroid},\mathrm{edge}\}$, the centered local coordinate is
\begin{equation}
\mathbf{u}^{\left(j,a\right)}\left(\mathbf{x}\right)
=
\mathbf{R}^{\left(j,a\right)}
\left(\mathbf{x}-\bar{\mathbf{x}}^{\left(j\right)}\right).
\end{equation}
The square side length is obtained from the larger of the two triangle extents in this local frame. With $\mathbf{u}_i^{\left(j,a\right)}=\mathbf{u}^{\left(j,a\right)}\left(\mathbf{v}_i^{\left(j\right)}\right)$ and expansion factor $\eta$, the half-width is
\begin{equation}
H^{\left(j,a\right)}
=
\frac{\eta}{2}
\max\left[
\max_i u_{i,x}^{\left(j,a\right)}-\min_i u_{i,x}^{\left(j,a\right)},
\max_i u_{i,y}^{\left(j,a\right)}-\min_i u_{i,y}^{\left(j,a\right)}
\right],
\end{equation}
and the spectral domain is
\begin{equation}
B^{\left(j,a\right)}
=
\left[-H^{\left(j,a\right)},H^{\left(j,a\right)}\right]
\times
\left[-H^{\left(j,a\right)},H^{\left(j,a\right)}\right].
\end{equation}
Thus $L_x^{\left(j,a\right)}=L_y^{\left(j,a\right)}=2H^{\left(j,a\right)}$. The coordinates passed to the NUFFT are shifted from centered coordinates to square-domain coordinates,
\begin{equation}
\left(x',y'\right)
=
\left(u_x^{\left(j,a\right)}+H^{\left(j,a\right)},u_y^{\left(j,a\right)}+H^{\left(j,a\right)}\right).
\end{equation}
This shift only changes the coordinate origin. The Fourier domain remains square. The active quadrature domain is selected by one of three masks inside this square domain,
\begin{align}
D^{\left(j,a\right)}
=
\begin{cases}
B^{\left(j,a\right)}, & \mathrm{square},\\
B^{\left(j,a\right)}\cap \left\{\mathbf{u}^{\left(j,a\right)}\left(\mathbf{x}\right):\mathbf{x}\in T^{\left(j\right)}\right\}, & \mathrm{triangle},\\
B^{\left(j,a\right)}\cap \left\{\mathbf{u}^{\left(j,a\right)}\left(\mathbf{x}\right):\left\|\mathbf{u}^{\left(j,a\right)}\right\|\leq H^{\left(j,a\right)}\right\}, & \mathrm{circle}.
\end{cases}
\end{align}
CSA is always evaluated on the square support respecting the local orientation.

\section{Quadrature Weights}\label{app:quadrature_weights}
The spectral coefficients require evaluation of the continuous integral
\begin{equation}
\hat{h}^{\left(j\right)}\left(\mathbf{k}\right) = \frac{1}{\left|D^{\left(j\right)}\right|}\int_{D^{\left(j\right)}} h\left(\mathbf{x}\right) e^{-i\mathbf{k}\cdot\mathbf{x}} \, d\mathbf{x},
\label{eq:continuous_integral}
\end{equation}
which must be approximated by a finite quadrature over the $Q^{\left(j\right)}$ scattered DEM samples $\left\{\mathbf{x}_q\right\}_{q=1}^{Q^{\left(j\right)}}$ inside the window as
\begin{equation}
\hat{h}^{\left(j\right)}\left(\mathbf{k}\right) \approx \frac{1}{\left|D^{\left(j\right)}\right|} \sum_{q=1}^{Q^{\left(j\right)}} w_q  h\left(\mathbf{x}_q\right) e^{-i\mathbf{k}\cdot\mathbf{x}_q}.
\label{eq:discrete_quadrature}
\end{equation}
The choice of weights $\left\{w_q\right\}$ determines the accuracy of this approximation for a given sampling geometry.
\subsection{Uniform Weights}
For near-uniform DEM sampling, where points are statistically indistinguishable from a homogeneous Poisson process, each point represents an equal share of the domain area. In that case, the weights reduce to
\begin{equation}
w_q = \frac{\left|D^{\left(j\right)}\right|}{Q^{\left(j\right)}}, \qquad q = 1,\dots,Q^{\left(j\right)},
\label{eq:uniform_weights}
\end{equation}
where $\left|D^{\left(j\right)}\right|$ is the area of the analysis domain, so the normalized coefficient multiplier is $w_q/\left|D^{\left(j\right)}\right| = 1/Q^{\left(j\right)}$. The resulting estimator is the Monte Carlo quadrature discussed in \cref{app:monte_carlo}, with the classical $\mathcal{O}\left[\left(Q^{\left(j\right)}\right)^{-1/2}\right]$ convergence rate \citep{Du_1999}. This scheme performs poorly when the sampling density varies appreciably across $D^{\left(j\right)}$, because clustered points are overcounted relative to sparsely sampled regions.

\subsection{Voronoi Weights}
For strongly non-uniform DEM distributions, e.g., along coastlines, a better approach is to weight each sample by the area of its geometric region of influence, obtained from the Voronoi tessellation of the point set \citep{Okabe_2000,Aurenhammer_1991}. The same construction underpins Voronoi-based natural-neighbor interpolation used on unstructured geophysical grids \citep{Braun_1995} and centroidal Voronoi quadrature \citep{Du_1999}. Given the points $P^{\left(j\right)} = \left\{\mathbf{x}_q\right\}_{q=1}^{Q^{\left(j\right)}}$ in $D^{\left(j\right)}$, the Voronoi cell associated with $\mathbf{x}_q$ is
\begin{equation}
V_q = \left\{\mathbf{x} : \left\|\mathbf{x}-\mathbf{x}_q\right\| \leq \left\|\mathbf{x}-\mathbf{x}_{q'}\right\|\ \forall q'\neq q\right\}
\label{eq:voronoi_cell}
\end{equation}
i.e., the set of locations closer to $\mathbf{x}_q$ than to any other sample. Each $V_q$ is a convex (possibly unbounded) polygon whose edges are segments of the perpendicular bisectors between $\mathbf{x}_q$ and its neighbors, and the collection $\left\{V_q\right\}$ tiles the plane. Because unbounded cells and cells extending past the window would contribute infinite or spurious area to the quadrature, every cell is intersected with the analysis domain, yielding
\begin{equation}
\tilde{V}_q^{\left(j\right)} = V_q \cap D^{\left(j\right)}.
\label{eq:clipped_cell}
\end{equation}
Rather than constructing $\tilde{V}_q^{\left(j\right)}$ explicitly with the Sutherland--Hodgman algorithm \citep{Sutherland_1974} and integrating it via the shoelace formula, the implementation used here evaluates $\left|\tilde{V}_q^{\left(j\right)}\right|$ via raster sampling. This choice handles arbitrary window shapes (triangular, circular, axis-aligned rectangle, or rotated rectangle) through a single code path and remains robust when DEM samples are nearly collinear or coincident. A regular Cartesian grid of $N_g \times N_g$ test points is laid over the bounding domain of $D^{\left(j\right)}$ with the cell-centered coordinates
\begin{equation}
\mathbf{g}_{ab} = \left[\frac{L_x^{\left(j\right)}}{N_g}\left(a + \frac{1}{2}\right), \frac{L_y^{\left(j\right)}}{N_g}\left(b + \frac{1}{2}\right)\right], \qquad a = 0,\dots,N_g-1, \quad b = 0,\dots,N_g-1,
\label{eq:raster_grid}
\end{equation}
and the subset $\mathcal{G}^{\left(j\right)} = \left\{\mathbf{g}_{ab} \in D^{\left(j\right)}\right\}$ lying inside the window is retained by an analytical in/out test specific to the window type. Each retained test point is then assigned to its nearest DEM sample by a $k$-d tree query.
\begin{equation}
q^\star\left(\mathbf{g}\right) = \underset{q}{\mathrm{argmin}} \left\|\mathbf{g} - \mathbf{x}_q\right\|,
\label{eq:nn_assignment}
\end{equation}
which is exactly the Voronoi membership test implied by \cref{eq:voronoi_cell}. Let $n_q$ be the number of test points in $\mathcal{G}^{\left(j\right)}$ that are assigned to DEM sample $\mathbf{x}_q$. The clipped-cell area is then estimated by scaling this count by the area represented by a single grid point, yielding
\begin{equation}
A_q^{\mathrm{Vor}}   \approx   n_q \cdot \frac{\left|D^{\left(j\right)}\right|}{\left|\mathcal{G}^{\left(j\right)}\right|},
\label{eq:raster_area}
\end{equation}
so that, by construction, the cell areas partition the window, implying
\begin{equation}
\sum_{q=1}^{Q^{\left(j\right)}} A_q^{\mathrm{Vor}}   =   \left|D^{\left(j\right)}\right|.
\label{eq:area_partition}
\end{equation}
\Cref{eq:raster_area} is a deterministic, low-discrepancy Monte Carlo estimate of $\left|\tilde{V}_q^{\left(j\right)}\right|$ whose error against the exact geometric area decays as $\mathcal{O}\left(N_g^{-1}\right)$. Empty cells ($n_q = 0$, which can occur when an isolated DEM point falls between grid lines) are assigned a small floor weight, and the full set is renormalized so that \cref{eq:area_partition} is preserved. The resulting quadrature in \cref{eq:discrete_quadrature} is a piecewise-constant Riemann sum on the rasterized Voronoi tessellation in which $h$ is approximated by its value at $\mathbf{x}_q$ on all of $\tilde{V}_q^{\left(j\right)}$. The Voronoi quadrature weights are therefore the clipped cell areas themselves.
\begin{equation}
w_q   =   A_q^{\mathrm{Vor}}, \qquad
\sum_{q=1}^{Q^{\left(j\right)}} w_q   =   \left|D^{\left(j\right)}\right|.
\label{eq:voronoi_weights}
\end{equation}
The dimensionless factor that enters the normalized coefficient is $w_q/\left|D^{\left(j\right)}\right| = A_q^{\mathrm{Vor}}/\left|D^{\left(j\right)}\right|$, so the coefficient contributions sum to unity without redefining $w_q$ as a probability weight. Points in dense clusters receive small cells and hence small contributions, while isolated points in sparse regions receive large cells, automatically correcting the density bias of \cref{eq:uniform_weights}. For smoothly sampled fields, the resulting quadrature error decays faster than the Monte Carlo rate and approaches the $\mathcal{O}\left[\left(Q^{\left(j\right)}\right)^{-1}\right]$ behavior of a locally constant trapezoidal rule on a centroidal Voronoi tessellation \citep{Du_1999}.

\section{Discrete Spectral Fidelity}\label{app:discrete_spectral_fidelity}
This section discusses the three main errors in the computed Fourier coefficients. These are Monte Carlo quadrature error due to finite sampling, spectral leakage caused by irregular sampling, and the Gibbs phenomenon arising from non-rectangular window boundaries.
\subsection{Monte Carlo Quadrature Error}\label{app:monte_carlo}
This subsection discusses the error due to finite sampling. Any discrete Fourier transform may approximate the continuous Fourier integral \cref{eq:cont_fourier} by a discrete sum over $Q$ randomly distributed samples \cref{eq:nufft_sum}. This is a Monte Carlo quadrature approximation to a continuous integral with uniform weights $w_q = \left|D\right|/Q$, where $\left|D\right|=L_xL_y$. Consider the monochromatic orography
\begin{equation}\label{eq:mono_monte_2}
h_q = A\cos\left(k_m x_q + \ell_n y_q\right) = \frac{A}{2}\left[e^{i \left(k_m x_q + \ell_n y_q\right)} + e^{-i\left(k_m x_q + \ell_n y_q\right)}\right].
\end{equation}
Note that the indices $\left(m,n\right)$ in \cref{eq:mono_monte_2} are not the same as those in \cref{eq:nufft_sum}. Substituting into \cref{eq:nufft_sum} and using linearity, one obtains
\begin{equation}
\hat h_{p,q} = \frac{A}{2Q}\sum_{r=1}^{Q}\left[e^{i \left(k_m - k_p\right) x_r + i \left(\ell_n - \ell_q\right) y_r} + e^{-i \left(k_m + k_p\right) x_r - i \left(\ell_n + \ell_q\right) y_r}\right],
\end{equation}
where $\left(p,q\right)$ are the mode indices of interest. For the case $\left(p,q\right)=\left(m,n\right)$, i.e., when the Fourier coefficients are evaluated at the true signal mode, the first exponential in the sum is $e^{0i}=1$, which implies
\begin{equation}
\hat h_{m,n} = \frac{A}{2Q}\sum_{r=1}^{Q}1 + \frac{A}{2Q}\sum_{r=1}^{Q}e^{-i \left(2k_m x_r + 2\ell_n y_r\right)} = \frac{A}{2} + \frac{A}{2Q}\sum_{r=1}^{Q} e^{-i \left(2k_m x_r + 2\ell_n y_r\right)}.
\end{equation}
The first term is the correct analytical value. The second term represents a discretization error arising from finite random sampling. This error term is an approximation of the continuous integral
\begin{equation}
\frac{A}{2\left|D\right|}\int_{D}e^{-i\left(2k_m x + 2\ell_n y\right)} \, dx \, dy.
\end{equation}
Over the periodic domain $D = \left[0,L_x\right]\times\left[0,L_y\right]$ with wavenumbers $k_m = 2\pi m/L_x$ and $\ell_n = 2\pi n/L_y$, this integral separates as
\begin{equation}
\frac{A}{2L_xL_y}\left[\int_{0}^{L_x}e^{-2ik_m x} \,dx\right]\left[\int_{0}^{L_y}e^{-2i\ell_n y} \, dy\right].
\end{equation}
The first of these integrals evaluates to
\begin{equation}
\int_{0}^{L_x}e^{-4\pi i m x/L_x} \, dx = \left[\frac{L_x}{-4\pi i m}e^{-4 \pi i m x/L_x}\right]_{0}^{L_x} = \frac{L_x}{-4 \pi i m}\left(e^{-4\pi i m} - 1\right) = 0
\end{equation}
since $e^{-4\pi i m} = 1$ for any integer $m$. The same argument applies to the second integral. Thus, the full integral vanishes exactly as the oscillating exponential completes an integer number of periods over the domain. However, with only $Q$ randomly placed sampling points, this cancellation is imperfect. To quantify the error, one can define a random variable
\begin{equation}
\epsilon = \frac{A}{2Q}\sum_{r=1}^{Q}z_r \qquad\mathrm{where}\quad z_r = e^{-i\left(2k_m x_r + 2\ell_n y_r\right)}.
\end{equation}
The Monte Carlo quadrature error specifically quantifies finite-sampling error under a uniform and independent distribution. Spectral leakage from irregular sampling is treated in the next subsection. Therefore, in this analysis, each sample point $\left(x_r, y_r\right)$ is uniformly distributed over the domain $D$. The expected value of each $z_r$ is
\begin{equation}
\mathbb{E}\left[z_r\right] = \frac{1}{\left|D\right|}\int_D e^{-i\left(2k_m x + 2\ell_n y\right)} \, dx  \, dy = 0
\end{equation}
by the preceding argument. Since $\left|z_r\right| = 1$ for all $r$ (complex exponentials have unit magnitude), the variance of each sample reads
\begin{equation}
\mathrm{Var}\left(z_r\right) = \mathbb{E}\left[\left|z_r\right|^2\right] - \left|\mathbb{E}\left[z_r\right]\right|^2 = 1 - 0 = 1.
\end{equation}
Since the samples $\left\{z_r\right\}_{r=1}^{Q}$ are independent and identically distributed, the variance of their sum is
\begin{equation}
\mathrm{Var}\left(\sum_{r=1}^{Q}z_r\right) = \sum_{r=1}^{Q}\mathrm{Var}\left(z_r\right) = Q,
\end{equation}
and using $\mathrm{Var}\left(cX\right)=c^2\mathrm{Var}\left(X\right)$ gives the variance of the error term as
\begin{equation}
\mathrm{Var}\left(\epsilon\right) = \mathrm{Var}\left(\frac{A}{2Q}\sum_{r=1}^{Q}z_r\right) = \left(\frac{A}{2Q}\right)^2\mathrm{Var}\left(\sum_{r=1}^{Q}z_r\right) = \frac{A^2}{4Q^2}\cdot Q = \frac{A^2}{4Q}.
\end{equation}
This establishes that the variance scales as $\mathcal{O}\left(Q^{-1}\right)$. The standard deviation, which characterizes the typical magnitude of the error, is therefore given by
\begin{equation}
\sigma_\epsilon = \sqrt{\mathrm{Var}\left(\epsilon\right)} = \frac{A}{2\sqrt{Q}},
\end{equation}
which demonstrates the characteristic $\mathcal{O}\left(Q^{-1/2}\right)$ convergence rate of Monte Carlo quadrature \citep{caflisch1998montecarloan}.

\subsection{Spectral Leakage Under Irregular Sampling}\label{app:spectral_leakage}
This subsection discusses spectral leakage caused by irregular sampling, distinguishing between uniform and non-uniform sampling density. On a uniform grid with regular spacing, the discrete Fourier transform recovers a monochromatic mode exactly because the basis functions
$e^{i\left(k_p x_r + \ell_q y_s\right)}$ are orthogonal in the discrete sum
\begin{equation}
\frac{1}{N_xN_y}\sum_{r=0}^{N_x-1}\sum_{s=0}^{N_y-1}
e^{i\left(k_p x_r + \ell_q y_s\right)}
e^{-i\left(k_{p'} x_r + \ell_{q'} y_s\right)}
= \delta_{pp'}\delta_{qq'}
\end{equation}
with $x_r=rL_x/N_x$ and $y_s=sL_y/N_y$. This orthogonality arises because the regular spacing generates the uniformly distributed phase samples $2\pi\left(p-p'\right)r/N_x$, which form a geometric series that sums to zero for $\left(p,q\right)\neq\left(p',q'\right)$. When sampling points are irregularly distributed, this discrete orthogonality is destroyed \citep{Coles2011CorrelatedNoise}. For general points $\left\{x_r,y_r\right\}$, the discrete inner product becomes
\begin{equation}
C_{pp'qq'} =
\frac{1}{Q}\sum_{r=1}^{Q}
e^{i\left(k_p-k_{p'}\right)x_r+i\left(\ell_q-\ell_{q'}\right)y_r},
\end{equation}
which is nonzero for $\left(p,q\right)\neq\left(p',q'\right)$. As a result, energy from the true mode leaks into other Fourier modes. If the sample locations are drawn independently from a uniform density $\rho\left(x,y\right)=\left|D\right|^{-1}$ over the domain, then
\begin{equation}
\mathbb{E}\left[C_{pp'qq'}\right]
= \frac{1}{\left|D\right|}\int_D
e^{i \left(\Delta k x+\Delta \ell y\right)} \, dx \, dy
= 0
\end{equation}
for all nonzero $\left(\Delta k,\Delta\ell\right)$.
Thus, discrete orthogonality holds only in expectation. Following the previous logic, the variance is
\begin{equation}
\mathrm{Var}\left(C_{pp'qq'}\right)=\frac{1}{Q},
\end{equation}
so the standard deviation scales as $\mathcal{O}\left(Q^{-1/2}\right)$. Multiplication by the signal amplitude $A/2$ yields typical spurious mode amplitudes of order $A/\left(2\sqrt{Q}\right)$. If instead the sampling points follow a non-uniform density $\rho\left(x,y\right)\neq \left|D\right|^{-1}$, one has
\begin{equation}
\mathbb{E}\left[C_{pp'qq'}\right]
= \int_D \rho\left(x,y\right)
e^{i\left(\Delta k x+\Delta \ell y\right)} \, dx \, dy
\neq 0.
\end{equation}
In this case, Fourier modes are not orthogonal even in expectation. \citet{PipeMenon1999SDC,Jackson1991Gridding} discuss this issue in detail, including practical mitigation strategies. In summary, irregular sampling always destroys exact discrete orthogonality. With uniform sampling density, this results in random, noise-like leakage that decreases slowly with sample size, whereas non-uniform sampling density introduces a spectral bias that remains even in the limit of infinitely many samples.

\subsection{Gibbs Phenomenon From Non-Rectangular Boundaries}\label{app:gibbs}

This subsection considers the Gibbs phenomenon arising from non-periodic mask boundaries. For the analysis domain $D^{\left(j\right)}$ restricted to a masked region, the orography field is effectively given by
\begin{equation}
h_{\mathrm{windowed}}\left(x,y\right) = \begin{cases} h\left(x,y\right) & \left(x,y\right) \in \mathrm{mask}, \\ 0 & \left(x,y\right) \notin \mathrm{mask}, \end{cases}
\end{equation}
and artificial discontinuities are introduced along the mask edges unless $h$ vanishes there. The resulting Gibbs overshoot is obtained from a one-dimensional section normal to a mask edge. Over a periodic interval of length $L_x$, a jump of magnitude $\Delta h$ is represented by
\begin{equation}
\label{eq:gibbs_step}
h_{\mathrm{windowed}}\left(x\right)
=
\begin{cases}
0, & -L_x/2 < x < 0,\\
\Delta h, & 0 < x < L_x/2.
\end{cases}
\end{equation}
With $k_m=2\pi m/L_x$, the Fourier coefficients are obtained by direct integration as
\begin{equation}
\begin{aligned}
\hat h_0
&=
\frac{1}{L_x}\int_{-L_x/2}^{L_x/2}
h_{\mathrm{windowed}}\left(x\right)\,\mathrm{d}x
=
\frac{\Delta h}{2},
\\
\hat h_m
&=
\frac{\Delta h}{L_x}\int_0^{L_x/2}e^{-ik_mx}\,\mathrm{d}x
=
\frac{\Delta h}{L_x}
\left[
\frac{e^{-ik_mx}}{-ik_m}
\right]_0^{L_x/2}
=
\frac{\Delta h}{2\pi i m}\left[1-\left(-1\right)^m\right],
\quad m\neq 0.
\end{aligned}
\label{eq:gibbs_coefficients}
\end{equation}
Since the even coefficients vanish, the truncation is indexed by $\mathcal{M}=2N+1$. By Euler's identity, the positive and negative modes are combined as
\begin{equation}
\hat h_m e^{ik_mx}
+
\hat h_{-m}e^{-ik_mx}
=
\frac{\Delta h}{\pi i m}
\left(
e^{ik_mx}-e^{-ik_mx}
\right)
=
\frac{2\Delta h}{\pi m}\sin\left(k_mx\right),
\quad m>0,\quad m\ \mathrm{odd}.
\label{eq:gibbs_mode_pair}
\end{equation}
The truncated series is therefore written as
\begin{equation}
S_{2N+1}\left(x\right)
=
\frac{\Delta h}{2}
+
\frac{2\Delta h}{\pi}
\sum_{r=0}^{N}
\frac{\sin\left(k_{2r+1}x\right)}{2r+1}.
\label{eq:gibbs_partial_sum}
\end{equation}
Since $k_{2r+1}=\left(2r+1\right)k_1$, the cosine sum required after differentiation is obtained directly as
\begin{equation}
\begin{aligned}
\sum_{r=0}^{N}\cos\left(k_{2r+1}x\right)
&=
\operatorname{Re}
\left[
e^{ik_1x}
\sum_{r=0}^{N}e^{2irk_1x}
\right]
=
\operatorname{Re}
\left[
e^{ik_1x}
\frac{1-e^{2i\left(N+1\right)k_1x}}
{1-e^{2ik_1x}}
\right]
\\
&=
\frac{\sin\left[2\left(N+1\right)k_1x\right]}
{2\sin\left(k_1x\right)}.
\end{aligned}
\label{eq:gibbs_cosine_sum}
\end{equation}
Consequently, the derivative of \cref{eq:gibbs_partial_sum} is obtained as
\begin{equation}
\begin{aligned}
\frac{\mathrm{d}S_{2N+1}}{\mathrm{d}x}
&=
\frac{4\Delta h}{L_x}
\sum_{r=0}^{N}
\cos\left(k_{2r+1}x\right)
\\
&=
\frac{2\Delta h}{L_x}
\frac{\sin\left[4\pi\left(N+1\right)x/L_x\right]}
{\sin\left(2\pi x/L_x\right)}.
\end{aligned}
\label{eq:gibbs_derivative}
\end{equation}
The first maximum on the positive side of the discontinuity therefore lies at $x_N=L_x/\left[4\left(N+1\right)\right]$. At this location, the midpoint Riemann-sum limit is
\begin{equation}
\begin{aligned}
\lim_{N\rightarrow\infty}
\left[
S_{2N+1}\left(x_N\right)-\Delta h
\right]
&=
\Delta h
\left[
\frac{1}{\pi}
\int_0^\pi\frac{\sin u}{u}\,\mathrm{d}u
-
\frac{1}{2}
\right]
\\
&\approx
0.08949\,\Delta h.
\end{aligned}
\label{eq:gibbs_overshoot}
\end{equation}
Thus, the limiting one-sided overshoot is approximately $9\,\%$ of the jump height \citep{David_1997,Hewitt_1979} and its distance from the discontinuity decreases as $\mathcal{O}\left(\mathcal{M}^{-1}\right)$. At the discontinuity, $S_{2N+1}\left(0\right)=\Delta h/2$. Unlike the errors in \cref{app:monte_carlo,app:spectral_leakage}, which decrease with $Q^{-1/2}$ under uniform sampling, the limiting Gibbs overshoot cannot be reduced by increasing $Q$.
\section{Shortcomings of $N_{\mathrm{eff}}$ and $S_{\delta}$}\label{app:ems_shortcomings}
Let $E_1 \geq E_2 \geq \cdots \geq E_{J^\star} \geq 0$ denote a non-increasing energy spectrum with $S = \sum_{j=1}^{J^\star} E_j > 0$. Fix a cap $J^\star=K_{\max}$ and a scale $\delta > 0$ such that one has
\begin{align}
    N_{\mathrm{eff}} &= \frac{\left(\sum_{j=1}^{J^\star} E_j\right)^2}
                   {\sum_{j=1}^{J^\star} E_j^2} \label{eq:neff},\\
    G_j   & = \frac{E_j}{E_{j+1}}, \qquad j=1,\dots,K_{\max}-1,\\
    S_{\delta} &= \frac{1}{K_{\max}-1}\sum_{j=1}^{K_{\max}-1}
              \exp  \left(-\frac{G_j-1}{\delta}\right) \label{eq:sdel}.
\end{align}
The following analysis shows that neither $N_{\mathrm{eff}}$ nor $S_{\delta}$ is adequate on its own. Each admits a family of spectra that it cannot correctly identify, and the two failure modes are essentially disjoint.

\subsection{$S_{\delta}$ Flat-Tail Degeneracy}
\label{app:sdel_blind}

$S_{\delta}$ can be driven arbitrarily close to $1$ by spectra that essentially concentrate all energy in a single mode. For some $\varepsilon \ll 1$, consider
\begin{equation}
    E^{\left(\varepsilon\right)} = \left(1, \varepsilon, \varepsilon, \dots, \varepsilon\right)
    \in \mathbb{R}^{K_{\max}}.
\end{equation}
The gap ratios are $G_1 = 1/\varepsilon$ and
$G_2 = G_3 = \dots = G_{K_{\max}-1} = 1$, giving
\begin{equation}
    S_{\delta}\left(\varepsilon\right)
    = \frac{1}{K_{\max}-1}\left[\exp  \left(-\frac{1/\varepsilon-1}{\delta}\right)
                            + K_{\max}-2\right]
      \xrightarrow[\varepsilon \to 0^+]{}   \frac{K_{\max}-2}{K_{\max}-1}\approx 1.
\end{equation}
The construction has one dominant mode and a long, negligible tail. Any reasonable truncation rule should set $K^{\star}=1$. However, $S_{\delta}$ saturates to 1, meaning it cannot distinguish the ratio $\varepsilon/\varepsilon = 1$ (near-zero neighbors) from the ratio $1/1 = 1$ (equal leaders), and it would incorrectly identify $K^{\star} = K_{\max}$ as the optimal truncation point. This failure is structural. $S_{\delta}$ is scale-invariant on each neighbor pair, so it rewards flatness at any magnitude, including noise-level tails. Meanwhile, for the same family of spectra, $N_{\mathrm{eff}}$ evaluates to
\begin{equation}
    N_{\mathrm{eff}}\left(\varepsilon\right)
    = \frac{\left[1+\left(K_{\max}-1\right)\varepsilon\right]^2}{1+\left(K_{\max}-1\right)\varepsilon^2}
      \xrightarrow[\varepsilon \to 0^+]{}   1,
\end{equation}
implying $\min\left(N_{\mathrm{eff}},K_{\max}\right)/K_{\max} \to 1/K_{\max}$. $N_{\mathrm{eff}}$ thus correctly identifies the single dominant mode and is unaffected by the long tail of small modes, which is the opposite behavior to that of $S_{\delta}$.
\subsection{$N_{\mathrm{eff}}$ Moment Degeneracy}
\label{app:neff_blind}

$N_{\mathrm{eff}}$ essentially compresses the whole spectrum to only two numbers,
\begin{equation}
    M_1   =   \sum_j E_j, \qquad
    M_2   =   \sum_j E_j^2,
\end{equation}
so that $N_{\mathrm{eff}}=M_1^2/M_2$. If two non-increasing spectra have the same pair $\left(M_1,M_2\right)$, then they have the same $N_{\mathrm{eff}}$, even if one is smooth and the other contains a sharp cliff. To exhibit the failure concretely, two very different spectra with identical $\left(M_1,M_2\right)$ are sufficient. For the first spectrum, consider the geometric decay
\begin{equation}
    E^A_j   =   \rho^{j-1}, \qquad j=1,\dots,K_{\max}, \quad 0<\rho<1.
\end{equation}
Its first two moments are
\begin{equation}
    M_1   =   \sum_{j=1}^{K_{\max}} E^A_j, \qquad
    M_2   =   \sum_{j=1}^{K_{\max}} \left(E^A_j\right)^2.
\end{equation}
For the second spectrum, consider the step function,
\begin{equation}
    E^B   =   \left(\underbrace{b,\dots,b}_{p\ \mathrm{times}},
               \underbrace{c,\dots,c}_{r\ \mathrm{times}}\right),
    \qquad b>c>0
\end{equation}
with two fixed integers $p,r$ with $p\in[1, K_{\max})$ and $r=K_{\max}-p$ . The two moments of this spectrum match those of $E^A$ when
\begin{equation}
    p b + r c   =   M_1, \qquad
    p b^2 + r c^2   =   M_2.
    \label{eq:matchshort}
\end{equation}
Manipulating the two equations in \eqref{eq:matchshort} then yields,
\begin{equation}
    K_{\max} M_2 - M_1^2
    = \left(p+r\right)\left(pb^2+rc^2\right) - \left(pb+rc\right)^2
    = p r \left(b-c\right)^2.
    \label{eq:momentgap}
\end{equation}
Hence $\left|b-c\right|$ is fixed by the moments. Since $E^A$ is not constant, it can be shown that the variance of $E^A$ is strictly positive, i.e., $K_{\max} M_2 - M_1^2 > 0$. Therefore,
\begin{equation}
    d   =   \sqrt{\frac{K_{\max} M_2 - M_1^2}{p r}}
    \label{eq:dformula}
\end{equation}
is real and strictly positive and the system \cref{eq:momentgap} has a unique solution, namely,
\begin{equation}
    b   =   \frac{M_1}{K_{\max}} + \frac{r}{K_{\max}}d, \qquad
    c   =   \frac{M_1}{K_{\max}} - \frac{p}{K_{\max}}d.
    \label{eq:bcformula}
\end{equation}
It remains to check positivity. The formula for $c$ yields
\begin{equation}
    c>0
    \iff M_1>p d
    \iff M_1^2 > p^2 d^2
    \iff M_1^2 > \frac{p\left(K_{\max} M_2-M_1^2\right)}{r}.
\end{equation}
Because of $p+r=K_{\max}$, this rearranges to
\begin{equation}
    M_1^2 > pM_2
    \iff \frac{M_1^2}{M_2} > p
    \iff N_{\mathrm{eff}}^A > p,
\end{equation}
Therefore, as long as $N_{\mathrm{eff}}^A > p$, one has $b=c+d>c>0$ and thus by construction,
\begin{equation}
    N_{\mathrm{eff}}^B   =   \frac{M_1^2}{M_2}   =   N_{\mathrm{eff}}^A.
\end{equation}
Meanwhile, in terms of gap ratio, for the geometric spectrum, one has
\begin{equation}
    G^A_j   =   \frac{E^A_j}{E^A_{j+1}}   =   \rho^{-1}
    \quad \forall j,
\end{equation}
whereas the two-level spectrum yields
\begin{equation}
    G^B_j   =
    \begin{cases}
        1, & j\neq p,\\
        b/c, & j=p.
    \end{cases}
\end{equation}
Therefore, for small $\delta$, one finds
\begin{equation}
    S_{\delta}^A   =   \exp  \left(-\frac{\rho^{-1}-1}{\delta}\right)\to0,
    \qquad
    S_{\delta}^B   \approx   \frac{K_{\max}-2}{K_{\max}-1}\to1.
\end{equation}
The first of these corresponds to a smooth decay and the second to a long plateau with one sharp drop, but $N_{\mathrm{eff}}$ assigns them the same value because it only registers $\left(M_1,M_2\right)$. In conclusion, $N_{\mathrm{eff}}$ is a moment functional, not a shape functional.

\section{Constrained Spectral Approximation}
\label{app:csam_comparison}
This section recasts the constrained spectral approximation method of \citet{chew2024aconstrained} in the notation of \cref{sec:elastic_mode_count} and \cref{sec:nufft}. It also expresses the final coefficient vectors in a common linear algebra form, demonstrating how the CSA coefficients differ from the ENUFFT coefficients defined by the quadrature sum \cref{eq:nufft_sum}.

\subsection{Common Basis Function}

On a local tangent-plane patch $D=\left[0,L_x\right]\times\left[0,L_y\right]$, let the orography be approximated by the truncated Fourier series
\begin{equation}
  h\left(x,y\right)
  \approx
  \sum_{m=-\mathcal{M}}^{\mathcal{M}}\sum_{n=-\mathcal{N}}^{\mathcal{N}}
  \hat h_{m,n}\,e^{\,i\left(k_m x + \ell_n y\right)},
  \qquad k_m=\frac{2\pi m}{L_x},\quad \ell_n=\frac{2\pi n}{L_y}.
  \label{eq:csam_series}
\end{equation}
The DEM is then given by the samples $\left\{x_q,y_q,h_q\right\}_{q=1}^{Q}$ of $h$ in the patch. For notational convenience, the double index $\left(m,n\right)$ is flattened into the single mode index
\begin{equation}
  j \equiv \left(m,n\right),\qquad j = 1,\dots,J, \qquad J=\left(2\mathcal{M}+1\right)\left(2\mathcal{N}+1\right)
\end{equation}
and the complex basis function
\begin{equation}
  \mu_j\left(x,y\right)
  = \exp\,\left[i\left(k_m x + \ell_n y\right)\right],
  \qquad j\equiv\left(m,n\right)
\end{equation}
is defined such that
\begin{equation}
  h\left(x,y\right)\approx \sum_{j=1}^{J} \hat h_j\,\mu_j\left(x,y\right).
\end{equation}
The complex coefficient vector is then assembled as
\begin{equation}
  \hat{ h} = \left(\hat h_1,\dots,\hat h_J\right)^{\mathsf{T}}\in\mathbb{C}^{J}
\end{equation}
and the data vector
\begin{equation}
   h = \left(h_1,\dots,h_Q\right)^{\mathsf{T}}\in\mathbb{C}^{Q},
  \qquad h_q = h\left(x_q,y_q\right)
\end{equation}
such that the link between coefficients and samples is given by the design matrix
\begin{equation}
  F \in \mathbb{C}^{Q\times J},
  \qquad
  F_{qj} = \mu_j\left(x_q,y_q\right)
  = \exp\left[i\left(k_m x_q + \ell_n y_q\right)\right],
  \label{eq:f_matrix_def}
\end{equation}
so that
\begin{equation}
  \left( F \hat{ h}\right)_q
  = \sum_{j=1}^{J} F_{qj}\,\hat h_j
  = \sum_{m,n} \hat h_{m,n}\,e^{i\left(k_m x_q + \ell_n y_q\right)}
\end{equation}
is the model prediction at the $q$th DEM point.

\subsection{CSA Coefficients}

\citet{chew2024aconstrained} define their coefficients as a regularized least-squares fit of the truncated Fourier expansion through an objective functional that reads
\begin{equation}
  J_{\mathrm{CSA}}\left(\hat{ h}\right)
  =
  \left\| F \hat{ h} -  h\right\|_2^2
  + \lambda\left\|\hat{ h}\right\|_2^2,
  \qquad \lambda\ge 0
  \label{eq:csam_objective}
\end{equation}
in the present notation, where $\lambda$ is a Tikhonov regularization parameter. The first term measures the mismatch between the reconstructed series and the DEM values in a discrete $L^2$ sense. The second term penalizes large coefficients and stabilizes the inversion in the presence of noisy or ill-conditioned data \citep{park2018parameterdet}. The minimizing coefficients are obtained by setting the gradient of \cref{eq:csam_objective} with respect to $\hat{ h}$ to zero. First, writing the objective as
\begin{equation}
  J_{\mathrm{CSA}}\left(\hat{ h}\right)
  = \left( F\hat{ h}- h\right)^{\ast}\left( F\hat{ h}- h\right)
  + \lambda\hat{ h}^{\ast}\hat{ h}
\end{equation}
and expanding yields
\begin{equation}
  J_{\mathrm{CSA}}\left(\hat{ h}\right)
  = \hat{ h}^{\ast} F^{\ast} F\hat{ h}
    - h^{\ast}F\hat h - \left(F\hat h\right)^{\ast}h
    +  h^{\ast} h
    + \lambda\,\hat{ h}^{\ast}\hat{ h}.
\end{equation}
Because of $\left(-h^{\ast}F\hat h\right)^{\ast}=-\left(F\hat h\right)^{\ast}h$ and since their sum is $-2\Re \left(\hat h^{\ast}F^{\ast}h\right)$, treating $\hat h, \hat h^{\ast}$ as independent and $\nabla_{\hat h}\Re \left(\hat h^{\ast}g\right)=g$, the complex gradient with respect to $\hat{ h}$ is
\begin{equation}
  \nabla_{\hat{ h}} J_{\mathrm{CSA}}
  = 2 F^{\ast} F\hat{ h}
    - 2 F^{\ast} h
    + 2\lambda \hat{ h}.
\end{equation}
Setting this gradient to zero gives the regularized normal equation
\begin{equation}
  \left( F^{\ast} F + \lambda I_J\right)\hat{ h}
  =  F^{\ast} h,
  \label{eq:csam_normal_eqs}
\end{equation}
where $ I_J$ is the $J\times J$ identity matrix. For $\lambda> 0$ and generic DEM sampling, the matrix $ F^{\ast} F + \lambda I_J$ is Hermitian [i.e., $\left(F^{\ast}F\right)^{\ast}=F^{\ast}F$], positive definite ($\left\|Fx\right\|_2^2 + \lambda \left\|x\right\|_2^2\ge \lambda \left\|x\right\|_2^2> 0$ for $x\neq 0$), and therefore invertible, leading to the unique CSA coefficient vector
\begin{equation}
  \hat{ h}^{\mathrm{CSA}}
  = \left( F^{\ast} F + \lambda I_J\right)^{-1} F^{\ast} h.
  \label{eq:csam_solution}
\end{equation}
\Cref{eq:csam_solution} collects the entire effect of the DEM sampling pattern, the choice of basis functions, and the regularization into the normal matrix $ F^{\ast} F + \lambda I_J$. In particular, for $\lambda=0$, it gives the ordinary least-squares (OLS) solution
\begin{equation}
  \hat{ h}^{\mathrm{CSA}}_{\lambda=0}
  = \left( F^{\ast} F\right)^{-1} F^{\ast} h,
\end{equation}
provided $ F^{\ast} F$ is non-singular.

\subsection{ENUFFT Coefficients}

In contrast, the ENUFFT approach starts from the continuous Fourier coefficients
\begin{equation}
  \hat h\left(k_m,\ell_n\right)
  = \frac{1}{\left|D\right|} \int_{0}^{L_x}\int_{0}^{L_y}
  h\left(x,y\right)\exp\left[-i\left(k_m x + \ell_n y\right)\right]\,dy\,dx
\end{equation}
and directly approximates this integral with the quadrature sum
\begin{equation}
  \hat h_{m,n}^{\mathrm{ENUFFT}}
  \approx \frac{1}{\left|D\right|}\sum_{q=1}^{Q} w_q h_q
  \exp\left[-i\left(k_m x_q + \ell_n y_q\right)\right].
  \label{eq:enufft_quadrature}
\end{equation}
For near-uniform dense DEM sampling, one has $w_q = \left|D\right|/Q$, which gives the discrete sum
\begin{equation}
  \hat h_{m,n}^{\mathrm{ENUFFT}}
  = \frac{1}{Q}\sum_{q=1}^{Q} h_q\,
  \exp\left[-i\left(k_m x_q + \ell_n y_q\right)\right].
\end{equation}
This is then evaluated efficiently with the NUFFT algorithm of \cref{sec:nufft}. In vector form, enumerating the modes with $j\equiv\left(m,n\right)$ as before, \cref{eq:enufft_quadrature} can be written as
\begin{equation}
  \hat{ h}^{\mathrm{ENUFFT}}
  =  C\, h,
  \qquad
  C_{j q}
  = \frac{w_q}{\left|D\right|}\exp\left[-i\left(k_m x_q + \ell_n y_q\right)\right].
  \label{eq:enufft_matrix}
\end{equation}
The coefficient matrix $ C$ depends only on the sampling locations and chosen quadrature weights. There is no matrix inversion and no regularization parameter. It is instructive to relate \cref{eq:enufft_matrix} to the CSA design matrix \cref{eq:f_matrix_def}. With the weights $w_q=\left|D\right|/Q$, the entries satisfy
\begin{equation}
  C_{j q}
  = \frac{1}{Q}\exp\left[-i\left(k_m x_q + \ell_n y_q\right)\right]
  = \frac{1}{Q}F_{qj}^*
\end{equation}
where $F_{qj}^*$ is the complex conjugate of $F_{qj}$, so that in matrix notation
\begin{equation}
  \hat{ h}^{\mathrm{ENUFFT}}
  = \frac{1}{Q}\, F^{\ast} h.
  \label{eq:enufft_vs_f}
\end{equation}
In summary, the CSA and ENUFFT solutions are given by
\begin{equation}
  \hat{ h}^{\mathrm{CSA}}
  = \left( F^{\ast} F + \lambda I_J\right)^{-1} F^{\ast} h,
  \qquad
  \hat{ h}^{\mathrm{ENUFFT}}
  = \frac{1}{Q} F^{\ast} h,
\end{equation}
i.e., the CSA coefficients result from applying the inverse normal matrix to $ F^{\ast} h$, whereas the ENUFFT coefficients are obtained directly from $ F^{\ast} h$ after scaling by $Q^{-1}$.

\subsection{Relation}

When the DEM lies on a strictly equidistant grid and periodic trapezoidal quadrature is exact for the Fourier modes under consideration, the columns of $ F$ are nearly orthogonal with respect to the discrete inner product. More precisely, for an $N_x\times N_y$ uniform grid, one finds
\begin{equation}
   F^{\ast} F \approx Q I_J,
\end{equation}
since $x$ and $y$ can be expressed as $L_xr/N_x$ and $L_ys/N_y$, respectively (with $Q=N_xN_y$), giving
\begin{equation}
  \left( F^{\ast} F\right)^{-1} F^{\ast} h
  \approx \frac{1}{Q} F^{\ast} h.
\end{equation}
In this idealized case, the ordinary-least-squares solution with $\lambda=0$ coincides with the ENUFFT coefficient vector up to small quadrature errors, implying
\begin{equation}
  \hat{ h}^{\mathrm{CSA}}_{\lambda=0}
  \approx \hat{ h}^{\mathrm{ENUFFT}}.
\end{equation}
On an irregular DEM, however, $ F^{\ast} F$ is no longer proportional to the identity matrix and its off-diagonal entries encode the non-orthogonality of the Fourier basis under the discrete sampling pattern. The CSA coefficients then become
\begin{equation}
  \hat{ h}^{\mathrm{CSA}}
  = \left( F^{\ast} F + \lambda I_J\right)^{-1} Q \hat{ h}^{\mathrm{ENUFFT}},
\end{equation}
which implies
\begin{equation}
  \hat{ h}^{\mathrm{CSA}}
  =  M^{-1}\hat{ h}^{\mathrm{ENUFFT}},
  \qquad
   M = \frac{1}{Q}\left( F^{\ast} F + \lambda I_J\right).
\end{equation}
Thus, the CSA coefficients can be interpreted as a filtered version of the quadrature-based ENUFFT coefficients, with the filter given by the inverse normal matrix.
\subsection{Computational Complexity}\label{app:csam_enufft_complexity}
In CSA, the core operation is solving the regularized normal equations
\begin{equation}
  \left( F^{\ast} F + \lambda I_J\right)\hat{ h}
  =  F^{\ast} h
  \label{eq:csam_normal_repeat}
\end{equation}
for each patch. Here $F\in\mathbb{C}^{Q\times J}$ has $Q$ DEM samples and $J$ Fourier modes, with $J=\left(2\mathcal{M}+1\right)\left(2\mathcal{N}+1\right)$. A standard dense least-squares solution based on normal equations proceeds by first forming the $J\times J$ normal matrix $F^{\ast}F$ and the right-hand side $F^{\ast}h$, and then factorizing the Hermitian matrix
\begin{equation}
  A_\lambda = F^{\ast}F + \lambda I_J.
\end{equation}
The cost of forming $F^{\ast}F$ is $\mathcal{O}\left(QJ^2\right)$, since each of the $J^2$ entries involves a sum over $Q$ DEM points, and the cost of a dense factorization of $A_\lambda$ is $\mathcal{O}\left(J^3\right)$ \citep{golub2012matrixcomput}. Thus, the overall complexity per patch scales as
\begin{equation}
  \mathcal{O}\left(QJ^2 + J^3\right),
\end{equation}
which becomes rapidly expensive as the number of retained modes $J$ grows. If the spectral domain is roughly square with $N_x\sim N_y\sim N$, then one has $J\sim N^2$ and thus
\begin{equation}
  \mathrm{CSA\ cost} \approx\mathcal{O}\left(QN^4 + N^6\right).
\end{equation}
The ENUFFT computation has a different structure. As described in \cref{sec:nufft}, the forward coefficients are obtained by spreading from $Q$ DEM points to an oversampled auxiliary grid of size $N_x^{\mathrm{aux}}\times N_y^{\mathrm{aux}}$, applying a standard FFT, and then deconvolving and truncating. The three steps have costs $\mathcal{O}\left(Q\right)$, $\mathcal{O}\left[N_x^{\mathrm{aux}}N_y^{\mathrm{aux}}\log\left(N_x^{\mathrm{aux}}N_y^{\mathrm{aux}}\right)\right]$, and $\mathcal{O}\left(N_x^{\mathrm{aux}}N_y^{\mathrm{aux}}\right)$, respectively, so that under the present simplification, the cumulative cost is
\begin{equation}
  \mathrm{ENUFFT\ cost} \approx \mathcal{O}\left(Q + N^2 \log N\right)
\end{equation}
with $N_x^{\mathrm{aux}}\sim N_y^{\mathrm{aux}}\sim N$. Both CSA and ENUFFT scale linearly with the number of DEM points $Q$, but ENUFFT scales only nearly linearly with the number of spectral modes through the $N^2\log N$ term, whereas CSA scales cubically through the $J^3\sim N^6$ term.
\section{Alps Preprocessing}\label{app:alps_preprocessing}
The elevations of the SRTM DEMs are first clipped below $-500\,\mathrm{m}$ to remove spurious ocean and void-fill values, yielding
\begin{equation}
  h_{ij}^{\mathrm{clip}}
  =
  \max\left(h_{ij}^{\textrm{raw}},-500~{\rm m}\right).
  \label{eq:alps_elevation_clip}
\end{equation}
The clipped mosaic is then coarse-grained from 1 arc-second to 30 arc-second resolution by non-overlapping block averages. This reduces the data to a more tractable resolution while also suppressing the finest, noise-dominated scales. For a block size $B=30$,
\begin{equation}
  h_{I,J}^{\textrm{cg}}
  =
  \frac{1}{B^2}
  \sum_{i=BI}^{B\left(I+1\right)-1}
  \sum_{j=BJ}^{B\left(J+1\right)-1}
  h_{i,j}^{\textrm{clip}}.
  \label{eq:alps_block_average}
\end{equation}
After the coarse-graining, a second low-pass filter is applied to remove terrain variance below the shortest scales intended for the source spectrum. The target filter is the $5\,\mathrm{km}$ spectral $e$-fold smoother used in scale-aware orographic source preprocessing \citep[e.g.,][]{vanniekerk2021scaleaware}. In spectral space, this filter is given by
\begin{equation}
  \hat h^{\textrm{smooth}}\left(k,\ell\right)
  =
  \hat h^{\textrm{cg}}\left(k,\ell\right)
  \exp\left[-\left(\frac{K}{K_0}\right)^2\right],
  \qquad
  K=\sqrt{k^2+\ell^2},
  \quad
  K_0=\frac{2\pi}{\lambda_0}
  \label{eq:alps_spectral_smoother}
\end{equation}
with $\lambda_0=5\,\mathrm{km}$. The equivalent normalized physical-space Gaussian reads
\begin{equation}
  G_{\textrm{eq}}\left(\Delta x,\Delta y\right)
  =
  \frac{1}{2\pi\sigma_{\textrm{eq}}^2}
  \exp\left(
    -\frac{\Delta x^2+\Delta y^2}{2\sigma_{\textrm{eq}}^2}
  \right)
  \label{eq:alps_gaussian_kernel}
\end{equation}
and its Fourier transform is $\exp\left(-\sigma_{\textrm{eq}}^2K^2/2\right)$. Matching this response to \cref{eq:alps_spectral_smoother} gives
\begin{equation}
  \sigma_{\textrm{eq}}
  =
  \frac{\sqrt{2}}{K_0}
  =
  \frac{\lambda_0}{\sqrt{2}\pi}
  \approx 1.13\,\mathrm{km}.
  \label{eq:alps_gaussian_sigma}
\end{equation}
On the processed grid, this corresponds to the Gaussian widths $\sigma_x\approx1.7$ and $\sigma_y\approx1.2$ (in grid points). Finally, the smoothed DEM is deplaned on the analysis mesh by removing the mean elevation of each target triangle, yielding
\begin{equation}
  h'_q = h_q - \bar h_j,
  \qquad
  \bar h_j = \frac{1}{\left|D^{\left(j\right)}\right|}\sum_{q\in T^{\left(j\right)}} w_qh_q.
  \label{eq:alps_deplaning}
\end{equation}
This ensures that the mean elevation is not parameterized as a subgrid-scale feature, since it is already carried by the resolved orography. 

Without these preprocessing steps, fine-scale noise and detail in the raw DEM would dominate the spectrum and alias (fold) into the retained modes, leading to spurious results. Additional preprocessing steps are documented in the accompanying software archive and repository \citep{banerjee2026enufftsoftware,banerjee2026elasticnufftgithub}.

\section{Mountain Wave}\label{app:mountain_wave}
Consider a flow over an orography given by
\begin{equation}
  h\left(\mathbf{x}\right)=h_{\symup{b}}\left(\mathbf{x}\right) + h_{\symup{w}}\left(\mathbf{x}\right)
\end{equation}
where $\mathbf{x}=\left(x,y\right)$. Simultaneously, let the three-dimensional flow be given by
\begin{equation}
  \mathbf{v}=\mathbf{v}_b+\mathbf{v}_w, \quad \mathbf{v}_{b}=\left(\mathbf{u},w\right)_{b}
  \quad \mathbf{v}_{w}=\left(\mathbf{u},w\right)_{w}.
\end{equation}
The orography is the model bottom boundary and is impermeable, so there is no normal flow. Any such surface can be described by $S\left(\mathbf{x},z\right)=z-h\left(\mathbf{x}\right)=0$. The normal vector to this surface is then $\nabla S=\left(-\partial_xh,-\partial_yh,1\right)$. Applying impermeability $\mathbf{v}\cdot\nabla S=0$ gives
\begin{equation}\label{eq:mountain_w}
  w=\mathbf{u}\cdot\nabla_hh
\end{equation}
where $\nabla_h$ is the horizontal gradient. Furthermore, the resolved flow must also satisfy impermeability independently, i.e., $w_b=\mathbf{u}_b\cdot\nabla_hh_{\symup{b}}$. Additionally, the normal wave velocity through the resolved background orography is $W_w=\mathbf{v}_w\cdot\nabla\left(z-h_{\symup{b}}\right)=w_w-\mathbf{u}_w\cdot\nabla_h h_{\symup{b}}$. Using these relations in \cref{eq:mountain_w} and neglecting nonlinear perturbation terms gives
\begin{equation}
  W_w=\mathbf{u}_b\cdot\nabla_hh_{\symup{w}}
\end{equation}
which is the normal linear perturbation through the resolved orography. This is the standard linear lower boundary condition for small-amplitude mountain waves \citep{teixeira2014physics,plougonven2020guide}. If the wave (unresolved) orography is expanded into its spectral modes, i.e.,
\begin{equation}
  h_{\symup{w}}\left(\mathbf{x}\right)=\sum_{\mu}\hat{h}_{\mu}\exp{\left(i\mathbf{k}_{\mu}\cdot\mathbf{x}\right)}=\sum_{\mu}h_{\symup{w},\mu}
\end{equation}
where $\mathbf{k}_{\mu}=\left(k,\ell\right)$ is the horizontal wave vector. Here $h_{\symup{w}}$ is the complex representation of the unresolved field, and the physical height is its real part. With this convention, the perturbation relation gives
\begin{equation}\label{eq:mountain_stationary_w}
  W_w=\sum_{\mu}i\mathbf{u}_b\cdot\mathbf{k}_{\mu}\hat{h}_{\mu}\exp{\left(i\mathbf{k}_{\mu}\cdot\mathbf{x}\right)}=\sum_{\mu}W_{w,\mu}
\end{equation}
such that
\begin{equation}
  W_{w,\mu}=i\mathbf{u}_b\cdot\mathbf{k}_{\mu}h_{\symup{w},\mu}.
\end{equation}
\cref{eq:mountain_stationary_w} shows that the perturbation induced by the orography follows the same stationary forcing phase, i.e., no time dependence and extrinsic frequency $\omega=0$. The perturbation is localized to the orography in space. Under a general Doppler shift, an observer moving with velocity $\mathbf{c}$ samples the phase change relative to the stationary frame $\partial_t\theta$ as $d_t\hat{\theta}=\partial_t\theta+\mathbf{c}\cdot\nabla\theta$. Using the convention $\omega=-d_t\theta$ gives $
\hat{\omega}=\omega - \mathbf{c}\cdot\nabla\theta
$.
For air flowing over the orography, experiencing its stationary forcing ($\omega_{\mu}=0$), and carrying a perturbation ($\theta_{\mu}=\mathbf{k}_{\mu}\cdot\mathbf{x}$) with velocity $\mathbf{v}_b$, the intrinsic frequency, i.e., the frequency perceived by the background winds, is
\begin{equation}\label{eq:mountain_omega}
\hat{\omega}_{\mu}=-\mathbf{k}_{\mu}\cdot\mathbf{u}_b,
\end{equation}
which gives the net perturbation as
\begin{equation}\label{eq:mountain_w_exp}
  W_w=-\sum_{\mu}i\hat{\omega}_{\mu}\hat{h}_{\mu}\exp{\left(i\mathbf{k}_{\mu}\cdot\mathbf{x}\right)}.
\end{equation}
\Cref{eq:mountain_w_exp} shows that the terrain spectrum fixes only the horizontal phase $\left(\mathbf{k}_{\mu}\right)$. The vertical phase, represented by wavenumber $\left(m_{\mu}\right)$, is obtained by assuming a gravity-wave response at the lower boundary with phase $\left(\mathbf{k}_{\mu},m_{\mu}\right)$. Subsequently, as shown in \citep{Achatz2017,Achatz2023}, a general monochromatic gravity wave of mode $\mu$ and phase $\left(\mathbf{k}_{\mu},m_{\mu}\right)$ under the WKB assumption follows the dispersion relation
\begin{equation}
  \mathbf{k}_\mu^2\left(N^2-\hat{\omega}_\mu^2\right)
  =
  m_\mu^2\left(\hat{\omega}_\mu^2-f^2\right), \qquad f<\left|\hat{\omega}_{\mu}\right|<N,
  \label{eq:mw_dispersion}
\end{equation}
and has phase-averaged total energy
\begin{equation}
  E_\mu
  =
  \frac{\bar{\rho}}{4}
  \left(
  \left|\vec{u}_{w,\mu}\right|^2+\left|W_{w,\mu}\right|^2+\frac{\left|b_{w,\mu}\right|^2}{N^2}
  \right).
  \label{eq:mw_energy}
\end{equation}
Using the linearized equations of motion together with continuity, \cref{eq:mountain_omega,eq:mountain_w_exp,eq:mw_dispersion}, and the wave-action density definition $\symscr{A}_\mu=E_\mu/\hat{\omega}_\mu$ gives
\begin{equation}
  \symscr{A}_\mu
  =
  -\frac{\bar{\rho}}{2}
  \mathbf{k}_{\mu}\cdot\mathbf{u}_b
  \frac{\mathbf{k}_\mu^2+m_\mu^2}{\mathbf{k}_\mu^2}
  \left|h_{\symup{w},\mu}\right|^2
  \label{eq:mw_action}
\end{equation}
showing that gravity-wave action depends not only on the terrain $\left|h_{\symup{w},\mu}\right|^2$ but also on the flow, $\mathbf{k}_{\mu}\cdot\mathbf{u}_b$.

\section{Result Tables}\label{app:result_tables}
\begin{sidewaystable}
\caption{Monochromatic-test peak-direction error $\Delta\theta$ ($^\circ$). Entries are pooled median $\mathbin{\pm}$ standard deviation.}
\label{tab:mono_direction_sweep}
\centering
\begin{tabular}{llrrrr}
\hline
Sweep parameter & Value & Square & Triangle & Circle & CSA \\
\hline
\multirow{6}{*}{Triangle orientation $\theta_T$} & $0^\circ$ & $26.20\mathbin{\pm}19.71$ & $0.00\mathbin{\pm}2.32$ & $23.20\mathbin{\pm}16.70$ & $23.20\mathbin{\pm}22.16$ \\
 & $60^\circ$ & $63.43\mathbin{\pm}26.76$ & $0.00\mathbin{\pm}2.75$ & $32.66\mathbin{\pm}27.73$ & $10.30\mathbin{\pm}27.15$ \\
 & $120^\circ$ & $32.66\mathbin{\pm}21.27$ & $0.00\mathbin{\pm}2.34$ & $26.57\mathbin{\pm}22.35$ & $24.78\mathbin{\pm}19.79$ \\
 & $180^\circ$ & $26.20\mathbin{\pm}12.99$ & $0.00\mathbin{\pm}2.15$ & $24.78\mathbin{\pm}12.17$ & $0.00\mathbin{\pm}15.35$ \\
 & $240^\circ$ & $26.20\mathbin{\pm}11.87$ & $0.00\mathbin{\pm}2.65$ & $24.78\mathbin{\pm}12.74$ & $23.20\mathbin{\pm}16.49$ \\
 & $300^\circ$ & $26.20\mathbin{\pm}16.02$ & $0.00\mathbin{\pm}3.56$ & $26.20\mathbin{\pm}15.87$ & $15.26\mathbin{\pm}14.50$ \\
\cline{2-6}
\multirow{3}{*}{Vertical offset $\Delta y/L$} & $-0.211$ & $26.20\mathbin{\pm}14.42$ & $0.00\mathbin{\pm}3.48$ & $26.20\mathbin{\pm}14.84$ & $17.92\mathbin{\pm}19.26$ \\
 & $-0.077$ & $26.20\mathbin{\pm}18.19$ & $0.00\mathbin{\pm}2.17$ & $24.78\mathbin{\pm}18.28$ & $18.43\mathbin{\pm}19.34$ \\
 & $0$ & $32.66\mathbin{\pm}24.57$ & $0.00\mathbin{\pm}2.13$ & $26.57\mathbin{\pm}24.08$ & $21.80\mathbin{\pm}21.59$ \\
\cline{2-6}
\multirow{3}{*}{Triangle uniformity $u$} & $0$ & $32.66\mathbin{\pm}18.40$ & $0.00\mathbin{\pm}4.40$ & $32.66\mathbin{\pm}18.76$ & $24.78\mathbin{\pm}21.82$ \\
 & $0.5$ & $26.57\mathbin{\pm}19.81$ & $0.00\mathbin{\pm}0.87$ & $24.78\mathbin{\pm}18.36$ & $18.43\mathbin{\pm}19.45$ \\
 & $1$ & $24.78\mathbin{\pm}18.00$ & $0.00\mathbin{\pm}0.00$ & $23.20\mathbin{\pm}16.32$ & $0.00\mathbin{\pm}16.90$ \\
\cline{2-6}
\multirow{3}{*}{Expansion ratio $r_{\rm exp}$} & $1$ & $24.78\mathbin{\pm}22.50$ & $0.00\mathbin{\pm}0.32$ & $0.00\mathbin{\pm}21.16$ & $0.00\mathbin{\pm}12.08$ \\
 & $1.5$ & $26.57\mathbin{\pm}17.53$ & $0.00\mathbin{\pm}1.78$ & $26.57\mathbin{\pm}16.61$ & $23.20\mathbin{\pm}20.55$ \\
 & $2$ & $26.57\mathbin{\pm}17.42$ & $0.00\mathbin{\pm}4.16$ & $26.57\mathbin{\pm}16.24$ & $26.57\mathbin{\pm}17.27$ \\
\cline{2-6}
\multirow{3}{*}{DEM point count $Q$} & $500$ & $26.57\mathbin{\pm}20.29$ & $0.00\mathbin{\pm}3.68$ & $26.20\mathbin{\pm}20.01$ & $18.43\mathbin{\pm}20.20$ \\
 & $1000$ & $26.20\mathbin{\pm}19.74$ & $0.00\mathbin{\pm}2.10$ & $26.20\mathbin{\pm}19.31$ & $18.43\mathbin{\pm}20.66$ \\
 & $2000$ & $26.20\mathbin{\pm}20.17$ & $0.00\mathbin{\pm}1.79$ & $26.20\mathbin{\pm}19.57$ & $18.43\mathbin{\pm}19.51$ \\
\cline{2-6}
\multirow{3}{*}{Oversampling $\sigma$} & $1.25$ & $26.57\mathbin{\pm}20.08$ & $0.00\mathbin{\pm}2.66$ & $26.20\mathbin{\pm}19.64$ & $18.43\mathbin{\pm}20.14$ \\
 & $1.5$ & $26.57\mathbin{\pm}20.08$ & $0.00\mathbin{\pm}2.68$ & $26.20\mathbin{\pm}19.64$ & $18.43\mathbin{\pm}20.14$ \\
 & $2$ & $26.57\mathbin{\pm}20.08$ & $0.00\mathbin{\pm}2.68$ & $26.20\mathbin{\pm}19.64$ & $18.43\mathbin{\pm}20.14$ \\
\cline{2-6}
\multirow{4}{*}{Mode-pair budget $K_{\max}$} & $4$ & $26.57\mathbin{\pm}16.35$ & $0.00\mathbin{\pm}4.21$ & $26.57\mathbin{\pm}17.90$ & $26.57\mathbin{\pm}19.71$ \\
 & $6$ & $24.78\mathbin{\pm}21.39$ & $0.00\mathbin{\pm}2.19$ & $18.43\mathbin{\pm}20.63$ & $10.30\mathbin{\pm}20.59$ \\
 & $8$ & $23.20\mathbin{\pm}21.14$ & $0.00\mathbin{\pm}1.53$ & $23.20\mathbin{\pm}20.67$ & $16.86\mathbin{\pm}16.83$ \\
 & $10$ & $33.11\mathbin{\pm}20.13$ & $0.00\mathbin{\pm}1.76$ & $26.20\mathbin{\pm}18.54$ & $21.80\mathbin{\pm}20.63$ \\
\cline{2-6}
\multirow{2}{*}{Point weights $w_q$} & Uniform & $26.20\mathbin{\pm}21.58$ & $0.00\mathbin{\pm}2.74$ & $26.20\mathbin{\pm}20.80$ & $18.43\mathbin{\pm}20.14$ \\
 & Voronoi area & $26.57\mathbin{\pm}18.33$ & $0.00\mathbin{\pm}2.60$ & $26.20\mathbin{\pm}18.38$ & $18.43\mathbin{\pm}20.14$ \\
\hline
\end{tabular}
\end{sidewaystable}

\begin{sidewaystable}
\caption{Monochromatic-test peak-amplitude error $\Delta A$ (m). Entries are pooled median $\mathbin{\pm}$ standard deviation.}
\label{tab:mono_amplitude_sweep}
\centering
\begin{tabular}{llrrrr}
\hline
Sweep parameter & Value & Square & Triangle & Circle & CSA \\
\hline
\multirow{6}{*}{Triangle orientation $\theta_T$} & $0^\circ$ & $286.6\mathbin{\pm}48.6$ & $17.4\mathbin{\pm}24.9$ & $273.5\mathbin{\pm}53.2$ & $155.8\mathbin{\pm}1.29\times10^{4}$ \\
 & $60^\circ$ & $284.0\mathbin{\pm}30.3$ & $17.1\mathbin{\pm}24.0$ & $278.9\mathbin{\pm}35.4$ & $46.6\mathbin{\pm}1875.1$ \\
 & $120^\circ$ & $294.8\mathbin{\pm}32.0$ & $19.6\mathbin{\pm}24.1$ & $283.7\mathbin{\pm}35.5$ & $136.1\mathbin{\pm}1354.5$ \\
 & $180^\circ$ & $269.2\mathbin{\pm}33.9$ & $14.9\mathbin{\pm}24.7$ & $243.2\mathbin{\pm}41.2$ & $2.4\mathbin{\pm}2.53\times10^{4}$ \\
 & $240^\circ$ & $277.6\mathbin{\pm}39.3$ & $20.6\mathbin{\pm}30.3$ & $259.7\mathbin{\pm}38.0$ & $163.4\mathbin{\pm}5869.5$ \\
 & $300^\circ$ & $282.0\mathbin{\pm}33.0$ & $17.1\mathbin{\pm}29.7$ & $259.5\mathbin{\pm}39.9$ & $114.5\mathbin{\pm}1889.8$ \\
\cline{2-6}
\multirow{3}{*}{Vertical offset $\Delta y/L$} & $-0.211$ & $261.0\mathbin{\pm}37.7$ & $18.1\mathbin{\pm}30.6$ & $242.4\mathbin{\pm}39.6$ & $122.4\mathbin{\pm}1.47\times10^{4}$ \\
 & $-0.077$ & $291.1\mathbin{\pm}32.4$ & $17.0\mathbin{\pm}25.0$ & $273.9\mathbin{\pm}38.6$ & $102.0\mathbin{\pm}6855.8$ \\
 & $0$ & $297.1\mathbin{\pm}31.7$ & $17.6\mathbin{\pm}23.1$ & $287.2\mathbin{\pm}38.7$ & $154.4\mathbin{\pm}1.29\times10^{4}$ \\
\cline{2-6}
\multirow{3}{*}{Triangle uniformity $u$} & $0$ & $262.2\mathbin{\pm}36.0$ & $23.3\mathbin{\pm}33.8$ & $255.4\mathbin{\pm}40.8$ & $462.1\mathbin{\pm}2.02\times10^{4}$ \\
 & $0.5$ & $296.7\mathbin{\pm}35.3$ & $17.1\mathbin{\pm}22.0$ & $284.1\mathbin{\pm}38.0$ & $111.0\mathbin{\pm}2900.5$ \\
 & $1$ & $292.6\mathbin{\pm}30.1$ & $14.2\mathbin{\pm}19.4$ & $264.6\mathbin{\pm}44.2$ & $0.0\mathbin{\pm}736.2$ \\
\cline{2-6}
\multirow{3}{*}{Expansion ratio $r_{\rm exp}$} & $1$ & $286.0\mathbin{\pm}35.6$ & $11.5\mathbin{\pm}14.7$ & $259.8\mathbin{\pm}39.2$ & $0.0\mathbin{\pm}126.2$ \\
 & $1.5$ & $283.3\mathbin{\pm}38.8$ & $18.2\mathbin{\pm}22.5$ & $274.1\mathbin{\pm}44.6$ & $151.8\mathbin{\pm}1.33\times10^{4}$ \\
 & $2$ & $276.7\mathbin{\pm}37.3$ & $28.0\mathbin{\pm}34.1$ & $267.2\mathbin{\pm}42.8$ & $941.3\mathbin{\pm}1.56\times10^{4}$ \\
\cline{2-6}
\multirow{3}{*}{DEM point count $Q$} & $500$ & $282.7\mathbin{\pm}37.4$ & $27.0\mathbin{\pm}31.8$ & $268.7\mathbin{\pm}41.9$ & $158.3\mathbin{\pm}4588.2$ \\
 & $1000$ & $282.4\mathbin{\pm}38.3$ & $19.9\mathbin{\pm}24.6$ & $266.6\mathbin{\pm}43.8$ & $105.3\mathbin{\pm}1.32\times10^{4}$ \\
 & $2000$ & $280.8\mathbin{\pm}36.9$ & $11.2\mathbin{\pm}17.5$ & $264.9\mathbin{\pm}42.4$ & $84.6\mathbin{\pm}1.53\times10^{4}$ \\
\cline{2-6}
\multirow{3}{*}{Oversampling $\sigma$} & $1.25$ & $281.8\mathbin{\pm}37.5$ & $17.7\mathbin{\pm}26.5$ & $267.1\mathbin{\pm}42.7$ & $120.4\mathbin{\pm}1.20\times10^{4}$ \\
 & $1.5$ & $281.8\mathbin{\pm}37.5$ & $17.7\mathbin{\pm}26.5$ & $267.1\mathbin{\pm}42.7$ & $120.4\mathbin{\pm}1.20\times10^{4}$ \\
 & $2$ & $281.8\mathbin{\pm}37.5$ & $17.7\mathbin{\pm}26.5$ & $267.1\mathbin{\pm}42.7$ & $120.4\mathbin{\pm}1.20\times10^{4}$ \\
\cline{2-6}
\multirow{4}{*}{Mode-pair budget $K_{\max}$} & $4$ & $284.1\mathbin{\pm}39.8$ & $16.5\mathbin{\pm}30.2$ & $263.4\mathbin{\pm}41.0$ & $156.6\mathbin{\pm}2057.4$ \\
 & $6$ & $278.0\mathbin{\pm}39.1$ & $18.6\mathbin{\pm}24.2$ & $256.8\mathbin{\pm}44.5$ & $63.8\mathbin{\pm}1255.6$ \\
 & $8$ & $280.9\mathbin{\pm}34.7$ & $15.5\mathbin{\pm}23.8$ & $271.8\mathbin{\pm}41.8$ & $124.0\mathbin{\pm}3791.9$ \\
 & $10$ & $285.2\mathbin{\pm}35.9$ & $19.8\mathbin{\pm}27.0$ & $274.9\mathbin{\pm}40.9$ & $89.5\mathbin{\pm}2.32\times10^{4}$ \\
\cline{2-6}
\multirow{2}{*}{Point weights $w_q$} & Uniform & $286.5\mathbin{\pm}38.2$ & $18.6\mathbin{\pm}26.8$ & $269.0\mathbin{\pm}44.1$ & $120.4\mathbin{\pm}1.20\times10^{4}$ \\
 & Voronoi area & $278.8\mathbin{\pm}36.6$ & $16.3\mathbin{\pm}26.2$ & $265.3\mathbin{\pm}41.2$ & $120.4\mathbin{\pm}1.20\times10^{4}$ \\
\hline
\end{tabular}
\end{sidewaystable}

\begin{sidewaystable}
\caption{Monochromatic-test retained mode-pair fraction $K^\star/K_{\max}$. Entries are pooled median $\mathbin{\pm}$ standard deviation.}
\label{tab:mono_budget_sweep}
\centering
\begin{tabular}{llrrrr}
\hline
Sweep parameter & Value & Square & Triangle & Circle & CSA \\
\hline
\multirow{6}{*}{Triangle orientation $\theta_T$} & $0^\circ$ & $0.625\mathbin{\pm}0.256$ & $0.750\mathbin{\pm}0.293$ & $0.750\mathbin{\pm}0.257$ & $1.000\mathbin{\pm}0.000$ \\
 & $60^\circ$ & $0.600\mathbin{\pm}0.222$ & $0.750\mathbin{\pm}0.282$ & $0.750\mathbin{\pm}0.224$ & $1.000\mathbin{\pm}0.000$ \\
 & $120^\circ$ & $0.750\mathbin{\pm}0.219$ & $0.750\mathbin{\pm}0.286$ & $0.875\mathbin{\pm}0.211$ & $1.000\mathbin{\pm}0.000$ \\
 & $180^\circ$ & $0.500\mathbin{\pm}0.215$ & $0.750\mathbin{\pm}0.286$ & $0.500\mathbin{\pm}0.243$ & $1.000\mathbin{\pm}0.000$ \\
 & $240^\circ$ & $0.500\mathbin{\pm}0.228$ & $0.750\mathbin{\pm}0.287$ & $0.625\mathbin{\pm}0.227$ & $1.000\mathbin{\pm}0.000$ \\
 & $300^\circ$ & $0.625\mathbin{\pm}0.225$ & $0.750\mathbin{\pm}0.277$ & $0.700\mathbin{\pm}0.237$ & $1.000\mathbin{\pm}0.000$ \\
\cline{2-6}
\multirow{3}{*}{Vertical offset $\Delta y/L$} & $-0.211$ & $0.500\mathbin{\pm}0.217$ & $0.750\mathbin{\pm}0.282$ & $0.500\mathbin{\pm}0.237$ & $1.000\mathbin{\pm}0.000$ \\
 & $-0.077$ & $0.667\mathbin{\pm}0.224$ & $0.750\mathbin{\pm}0.287$ & $0.750\mathbin{\pm}0.231$ & $1.000\mathbin{\pm}0.000$ \\
 & $0$ & $0.667\mathbin{\pm}0.230$ & $0.750\mathbin{\pm}0.285$ & $0.833\mathbin{\pm}0.219$ & $1.000\mathbin{\pm}0.000$ \\
\cline{2-6}
\multirow{3}{*}{Triangle uniformity $u$} & $0$ & $0.500\mathbin{\pm}0.208$ & $1.000\mathbin{\pm}0.227$ & $0.500\mathbin{\pm}0.240$ & $1.000\mathbin{\pm}0.000$ \\
 & $0.5$ & $0.667\mathbin{\pm}0.224$ & $0.750\mathbin{\pm}0.269$ & $0.750\mathbin{\pm}0.229$ & $1.000\mathbin{\pm}0.000$ \\
 & $1$ & $0.700\mathbin{\pm}0.222$ & $0.600\mathbin{\pm}0.300$ & $0.750\mathbin{\pm}0.236$ & $1.000\mathbin{\pm}0.000$ \\
\cline{2-6}
\multirow{3}{*}{Expansion ratio $r_{\rm exp}$} & $1$ & $0.667\mathbin{\pm}0.235$ & $0.375\mathbin{\pm}0.186$ & $0.750\mathbin{\pm}0.233$ & $1.000\mathbin{\pm}0.000$ \\
 & $1.5$ & $0.600\mathbin{\pm}0.242$ & $0.750\mathbin{\pm}0.195$ & $0.750\mathbin{\pm}0.248$ & $1.000\mathbin{\pm}0.000$ \\
 & $2$ & $0.500\mathbin{\pm}0.219$ & $1.000\mathbin{\pm}0.102$ & $0.667\mathbin{\pm}0.244$ & $1.000\mathbin{\pm}0.000$ \\
\cline{2-6}
\multirow{3}{*}{DEM point count $Q$} & $500$ & $0.750\mathbin{\pm}0.218$ & $0.875\mathbin{\pm}0.257$ & $0.900\mathbin{\pm}0.213$ & $1.000\mathbin{\pm}0.000$ \\
 & $1000$ & $0.500\mathbin{\pm}0.226$ & $0.750\mathbin{\pm}0.287$ & $0.700\mathbin{\pm}0.233$ & $1.000\mathbin{\pm}0.000$ \\
 & $2000$ & $0.500\mathbin{\pm}0.218$ & $0.667\mathbin{\pm}0.294$ & $0.500\mathbin{\pm}0.232$ & $1.000\mathbin{\pm}0.000$ \\
\cline{2-6}
\multirow{3}{*}{Oversampling $\sigma$} & $1.25$ & $0.600\mathbin{\pm}0.235$ & $0.750\mathbin{\pm}0.285$ & $0.750\mathbin{\pm}0.243$ & $1.000\mathbin{\pm}0.000$ \\
 & $1.5$ & $0.600\mathbin{\pm}0.235$ & $0.750\mathbin{\pm}0.285$ & $0.750\mathbin{\pm}0.243$ & $1.000\mathbin{\pm}0.000$ \\
 & $2$ & $0.600\mathbin{\pm}0.235$ & $0.750\mathbin{\pm}0.285$ & $0.750\mathbin{\pm}0.243$ & $1.000\mathbin{\pm}0.000$ \\
\cline{2-6}
\multirow{4}{*}{Mode-pair budget $K_{\max}$} & $4$ & $0.750\mathbin{\pm}0.220$ & $1.000\mathbin{\pm}0.261$ & $0.750\mathbin{\pm}0.209$ & $1.000\mathbin{\pm}0.000$ \\
 & $6$ & $0.500\mathbin{\pm}0.219$ & $0.833\mathbin{\pm}0.283$ & $0.500\mathbin{\pm}0.239$ & $1.000\mathbin{\pm}0.000$ \\
 & $8$ & $0.500\mathbin{\pm}0.230$ & $0.750\mathbin{\pm}0.289$ & $0.750\mathbin{\pm}0.246$ & $1.000\mathbin{\pm}0.000$ \\
 & $10$ & $0.600\mathbin{\pm}0.255$ & $0.700\mathbin{\pm}0.294$ & $0.800\mathbin{\pm}0.243$ & $1.000\mathbin{\pm}0.000$ \\
\cline{2-6}
\multirow{2}{*}{Point weights $w_q$} & Uniform & $0.667\mathbin{\pm}0.231$ & $0.800\mathbin{\pm}0.277$ & $0.750\mathbin{\pm}0.230$ & $1.000\mathbin{\pm}0.000$ \\
 & Voronoi area & $0.500\mathbin{\pm}0.227$ & $0.750\mathbin{\pm}0.291$ & $0.667\mathbin{\pm}0.248$ & $1.000\mathbin{\pm}0.000$ \\
\hline
\end{tabular}
\end{sidewaystable}

\begin{sidewaystable}
\caption{Alpine-test relative reconstructed RMSE $\varepsilon_{\rm rel}$ for R2B4 ($\Delta=160\,\mathrm{km}$). Entries are pooled median $\mathbin{\pm}$ standard deviation.}
\label{tab:alps_r2b4_rmse}
\centering
\begin{tabular}{llrrrr}
\hline
Sweep parameter & Value & Square & Triangle & Circle & CSA \\
\hline
\multirow{2}{*}{Window alignment} & Centroid & $0.808\mathbin{\pm}0.902$ & $1.635\mathbin{\pm}0.969$ & $0.797\mathbin{\pm}1.094$ & $0.583\mathbin{\pm}0.220$ \\
 & Edge aligned & $0.842\mathbin{\pm}1.168$ & $3.278\mathbin{\pm}1.743$ & $0.872\mathbin{\pm}1.476$ & $0.766\mathbin{\pm}0.227$ \\
\cline{2-6}
\multirow{2}{*}{Point weights $w_q$} & Uniform & $0.789\mathbin{\pm}1.043$ & $2.153\mathbin{\pm}1.602$ & $0.805\mathbin{\pm}1.402$ & $0.630\mathbin{\pm}0.230$ \\
 & Voronoi area & $0.878\mathbin{\pm}1.039$ & $2.183\mathbin{\pm}1.597$ & $0.870\mathbin{\pm}1.193$ & $0.630\mathbin{\pm}0.230$ \\
\cline{2-6}
\multirow{3}{*}{Expansion ratio $r_{\rm exp}$} & $1$ & $0.779\mathbin{\pm}0.769$ & $0.988\mathbin{\pm}0.330$ & $0.793\mathbin{\pm}0.885$ & $0.573\mathbin{\pm}0.220$ \\
 & $1.5$ & $0.843\mathbin{\pm}1.042$ & $2.250\mathbin{\pm}0.847$ & $0.837\mathbin{\pm}1.042$ & $0.674\mathbin{\pm}0.242$ \\
 & $2$ & $0.888\mathbin{\pm}1.241$ & $3.488\mathbin{\pm}1.367$ & $0.921\mathbin{\pm}1.768$ & $0.722\mathbin{\pm}0.221$ \\
\cline{2-6}
\multirow{3}{*}{Oversampling $\sigma$} & $1.25$ & $0.818\mathbin{\pm}1.045$ & $2.183\mathbin{\pm}1.600$ & $0.835\mathbin{\pm}1.304$ & $0.630\mathbin{\pm}0.230$ \\
 & $1.5$ & $0.818\mathbin{\pm}1.046$ & $2.183\mathbin{\pm}1.600$ & $0.835\mathbin{\pm}1.304$ & $0.630\mathbin{\pm}0.230$ \\
 & $2$ & $0.818\mathbin{\pm}1.046$ & $2.183\mathbin{\pm}1.600$ & $0.835\mathbin{\pm}1.304$ & $0.630\mathbin{\pm}0.230$ \\
\hline
\end{tabular}

\vspace{1.25em}
\caption{Alpine-test relative reconstructed RMSE $\varepsilon_{\rm rel}$ for R2B5 ($\Delta=80\,\mathrm{km}$). Entries are pooled median $\mathbin{\pm}$ standard deviation.}
\label{tab:alps_r2b5_rmse}
\centering
\begin{tabular}{llrrrr}
\hline
Sweep parameter & Value & Square & Triangle & Circle & CSA \\
\hline
\multirow{2}{*}{Window alignment} & Centroid & $0.801\mathbin{\pm}1.318$ & $1.736\mathbin{\pm}1.163$ & $0.784\mathbin{\pm}1.248$ & $0.326\mathbin{\pm}0.135$ \\
 & Edge aligned & $0.797\mathbin{\pm}1.615$ & $4.037\mathbin{\pm}2.385$ & $0.805\mathbin{\pm}1.357$ & $0.375\mathbin{\pm}0.208$ \\
\cline{2-6}
\multirow{2}{*}{Point weights $w_q$} & Uniform & $0.789\mathbin{\pm}1.414$ & $2.034\mathbin{\pm}2.168$ & $0.785\mathbin{\pm}1.173$ & $0.350\mathbin{\pm}0.179$ \\
 & Voronoi area & $0.810\mathbin{\pm}1.518$ & $2.200\mathbin{\pm}2.180$ & $0.806\mathbin{\pm}1.413$ & $0.350\mathbin{\pm}0.179$ \\
\cline{2-6}
\multirow{3}{*}{Expansion ratio $r_{\rm exp}$} & $1$ & $0.790\mathbin{\pm}0.763$ & $0.905\mathbin{\pm}0.410$ & $0.777\mathbin{\pm}0.607$ & $0.306\mathbin{\pm}0.096$ \\
 & $1.5$ & $0.788\mathbin{\pm}1.436$ & $2.212\mathbin{\pm}1.248$ & $0.787\mathbin{\pm}1.319$ & $0.356\mathbin{\pm}0.149$ \\
 & $2$ & $0.819\mathbin{\pm}1.952$ & $4.051\mathbin{\pm}1.960$ & $0.825\mathbin{\pm}1.710$ & $0.411\mathbin{\pm}0.234$ \\
\cline{2-6}
\multirow{3}{*}{Oversampling $\sigma$} & $1.25$ & $0.800\mathbin{\pm}1.475$ & $2.159\mathbin{\pm}2.176$ & $0.793\mathbin{\pm}1.306$ & $0.350\mathbin{\pm}0.179$ \\
 & $1.5$ & $0.800\mathbin{\pm}1.475$ & $2.159\mathbin{\pm}2.176$ & $0.793\mathbin{\pm}1.306$ & $0.350\mathbin{\pm}0.179$ \\
 & $2$ & $0.800\mathbin{\pm}1.475$ & $2.159\mathbin{\pm}2.176$ & $0.793\mathbin{\pm}1.306$ & $0.350\mathbin{\pm}0.179$ \\
\hline
\end{tabular}
\end{sidewaystable}

\begin{sidewaystable}
\caption{Alpine-test retained mode-pair fraction $K^\star/K_{\max}$ for R2B4 ($\Delta=160\,\mathrm{km}$). Entries are pooled median $\mathbin{\pm}$ standard deviation.}
\label{tab:alps_r2b4_budget}
\centering
\begin{tabular}{llrrrr}
\hline
Sweep parameter & Value & Square & Triangle & Circle & CSA \\
\hline
\multirow{2}{*}{Window alignment} & Centroid & $0.312\mathbin{\pm}0.315$ & $0.438\mathbin{\pm}0.309$ & $0.312\mathbin{\pm}0.285$ & $1.000\mathbin{\pm}0.000$ \\
 & Edge aligned & $0.562\mathbin{\pm}0.313$ & $0.750\mathbin{\pm}0.287$ & $0.562\mathbin{\pm}0.310$ & $1.000\mathbin{\pm}0.000$ \\
\cline{2-6}
\multirow{2}{*}{Point weights $w_q$} & Uniform & $0.375\mathbin{\pm}0.259$ & $0.562\mathbin{\pm}0.319$ & $0.375\mathbin{\pm}0.261$ & $1.000\mathbin{\pm}0.000$ \\
 & Voronoi area & $0.688\mathbin{\pm}0.346$ & $0.562\mathbin{\pm}0.318$ & $0.562\mathbin{\pm}0.333$ & $1.000\mathbin{\pm}0.000$ \\
\cline{2-6}
\multirow{3}{*}{Expansion ratio $r_{\rm exp}$} & $1$ & $0.250\mathbin{\pm}0.280$ & $0.250\mathbin{\pm}0.247$ & $0.312\mathbin{\pm}0.265$ & $1.000\mathbin{\pm}0.000$ \\
 & $1.5$ & $0.438\mathbin{\pm}0.323$ & $0.562\mathbin{\pm}0.266$ & $0.438\mathbin{\pm}0.305$ & $1.000\mathbin{\pm}0.000$ \\
 & $2$ & $0.688\mathbin{\pm}0.293$ & $1.000\mathbin{\pm}0.208$ & $0.688\mathbin{\pm}0.281$ & $1.000\mathbin{\pm}0.000$ \\
\cline{2-6}
\multirow{3}{*}{Oversampling $\sigma$} & $1.25$ & $0.438\mathbin{\pm}0.329$ & $0.562\mathbin{\pm}0.318$ & $0.438\mathbin{\pm}0.314$ & $1.000\mathbin{\pm}0.000$ \\
 & $1.5$ & $0.438\mathbin{\pm}0.329$ & $0.562\mathbin{\pm}0.318$ & $0.438\mathbin{\pm}0.314$ & $1.000\mathbin{\pm}0.000$ \\
 & $2$ & $0.438\mathbin{\pm}0.329$ & $0.562\mathbin{\pm}0.318$ & $0.438\mathbin{\pm}0.314$ & $1.000\mathbin{\pm}0.000$ \\
\hline
\end{tabular}

\vspace{1.25em}
\caption{Alpine-test retained mode-pair fraction $K^\star/K_{\max}$ for R2B5 ($\Delta=80\,\mathrm{km}$). Entries are pooled median $\mathbin{\pm}$ standard deviation.}
\label{tab:alps_r2b5_budget}
\centering
\begin{tabular}{llrrrr}
\hline
Sweep parameter & Value & Square & Triangle & Circle & CSA \\
\hline
\multirow{2}{*}{Window alignment} & Centroid & $0.156\mathbin{\pm}0.306$ & $0.281\mathbin{\pm}0.297$ & $0.188\mathbin{\pm}0.295$ & $1.000\mathbin{\pm}0.000$ \\
 & Edge aligned & $0.438\mathbin{\pm}0.348$ & $0.562\mathbin{\pm}0.322$ & $0.406\mathbin{\pm}0.348$ & $1.000\mathbin{\pm}0.000$ \\
\cline{2-6}
\multirow{2}{*}{Point weights $w_q$} & Uniform & $0.125\mathbin{\pm}0.230$ & $0.344\mathbin{\pm}0.333$ & $0.156\mathbin{\pm}0.241$ & $1.000\mathbin{\pm}0.000$ \\
 & Voronoi area & $0.469\mathbin{\pm}0.373$ & $0.438\mathbin{\pm}0.329$ & $0.438\mathbin{\pm}0.367$ & $1.000\mathbin{\pm}0.000$ \\
\cline{2-6}
\multirow{3}{*}{Expansion ratio $r_{\rm exp}$} & $1$ & $0.094\mathbin{\pm}0.296$ & $0.125\mathbin{\pm}0.231$ & $0.094\mathbin{\pm}0.273$ & $1.000\mathbin{\pm}0.000$ \\
 & $1.5$ & $0.312\mathbin{\pm}0.320$ & $0.406\mathbin{\pm}0.298$ & $0.312\mathbin{\pm}0.320$ & $1.000\mathbin{\pm}0.000$ \\
 & $2$ & $0.531\mathbin{\pm}0.325$ & $0.719\mathbin{\pm}0.279$ & $0.500\mathbin{\pm}0.324$ & $1.000\mathbin{\pm}0.000$ \\
\cline{2-6}
\multirow{3}{*}{Oversampling $\sigma$} & $1.25$ & $0.281\mathbin{\pm}0.343$ & $0.406\mathbin{\pm}0.333$ & $0.281\mathbin{\pm}0.336$ & $1.000\mathbin{\pm}0.000$ \\
 & $1.5$ & $0.281\mathbin{\pm}0.343$ & $0.406\mathbin{\pm}0.333$ & $0.281\mathbin{\pm}0.336$ & $1.000\mathbin{\pm}0.000$ \\
 & $2$ & $0.281\mathbin{\pm}0.343$ & $0.406\mathbin{\pm}0.333$ & $0.281\mathbin{\pm}0.336$ & $1.000\mathbin{\pm}0.000$ \\
\hline
\end{tabular}
\end{sidewaystable}

\begin{sidewaystable}
\caption{Alpine-test spectral-to-physical variance ratio $\sigma^2_{\rm spec}/\sigma^2_{\rm phys}$ for R2B4 ($\Delta=160\,\mathrm{km}$). Entries are pooled median $\mathbin{\pm}$ standard deviation.}
\label{tab:alps_r2b4_variance}
\centering
\begin{tabular}{llrrrr}
\hline
Sweep parameter & Value & Square & Triangle & Circle & CSA \\
\hline
\multirow{2}{*}{Window alignment} & Centroid & $0.53\mathbin{\pm}24.23$ & $1.76\mathbin{\pm}0.94$ & $0.56\mathbin{\pm}23.86$ & $4.82\mathbin{\pm}3.87\times10^{5}$ \\
 & Edge aligned & $0.68\mathbin{\pm}37.89$ & $3.05\mathbin{\pm}1.47$ & $0.76\mathbin{\pm}42.17$ & $24.06\mathbin{\pm}5.29\times10^{5}$ \\
\cline{2-6}
\multirow{2}{*}{Point weights $w_q$} & Uniform & $0.51\mathbin{\pm}27.75$ & $1.95\mathbin{\pm}1.39$ & $0.57\mathbin{\pm}31.62$ & $9.73\mathbin{\pm}4.63\times10^{5}$ \\
 & Voronoi area & $0.74\mathbin{\pm}35.42$ & $1.93\mathbin{\pm}1.39$ & $0.75\mathbin{\pm}36.78$ & $9.73\mathbin{\pm}4.63\times10^{5}$ \\
\cline{2-6}
\multirow{3}{*}{Expansion ratio $r_{\rm exp}$} & $1$ & $0.45\mathbin{\pm}12.15$ & $1.04\mathbin{\pm}0.44$ & $0.50\mathbin{\pm}7.32$ & $1.09\mathbin{\pm}643.00$ \\
 & $1.5$ & $0.64\mathbin{\pm}34.34$ & $1.99\mathbin{\pm}0.73$ & $0.70\mathbin{\pm}31.20$ & $16.23\mathbin{\pm}3.73\times10^{4}$ \\
 & $2$ & $0.75\mathbin{\pm}41.19$ & $3.23\mathbin{\pm}1.24$ & $0.82\mathbin{\pm}49.76$ & $277.39\mathbin{\pm}7.96\times10^{5}$ \\
\cline{2-6}
\multirow{3}{*}{Oversampling $\sigma$} & $1.25$ & $0.58\mathbin{\pm}31.84$ & $1.95\mathbin{\pm}1.39$ & $0.64\mathbin{\pm}34.31$ & $9.73\mathbin{\pm}4.63\times10^{5}$ \\
 & $1.5$ & $0.58\mathbin{\pm}31.84$ & $1.95\mathbin{\pm}1.39$ & $0.64\mathbin{\pm}34.31$ & $9.73\mathbin{\pm}4.63\times10^{5}$ \\
 & $2$ & $0.58\mathbin{\pm}31.84$ & $1.95\mathbin{\pm}1.39$ & $0.64\mathbin{\pm}34.31$ & $9.73\mathbin{\pm}4.63\times10^{5}$ \\
\hline
\end{tabular}

\vspace{1.25em}
\caption{Alpine-test spectral-to-physical variance ratio $\sigma^2_{\rm spec}/\sigma^2_{\rm phys}$ for R2B5 ($\Delta=80\,\mathrm{km}$). Entries are pooled median $\mathbin{\pm}$ standard deviation.}
\label{tab:alps_r2b5_variance}
\centering
\begin{tabular}{llrrrr}
\hline
Sweep parameter & Value & Square & Triangle & Circle & CSA \\
\hline
\multirow{2}{*}{Window alignment} & Centroid & $0.48\mathbin{\pm}26.79$ & $1.90\mathbin{\pm}1.16$ & $0.58\mathbin{\pm}24.49$ & $495.14\mathbin{\pm}1.41\times10^{6}$ \\
 & Edge aligned & $0.63\mathbin{\pm}183.04$ & $3.87\mathbin{\pm}2.00$ & $0.71\mathbin{\pm}77.93$ & $1.57\times10^{4}\mathbin{\pm}4.40\times10^{6}$ \\
\cline{2-6}
\multirow{2}{*}{Point weights $w_q$} & Uniform & $0.45\mathbin{\pm}107.35$ & $2.08\mathbin{\pm}1.90$ & $0.55\mathbin{\pm}55.08$ & $2637.79\mathbin{\pm}3.35\times10^{6}$ \\
 & Voronoi area & $0.76\mathbin{\pm}150.81$ & $2.37\mathbin{\pm}1.89$ & $0.80\mathbin{\pm}60.45$ & $2637.79\mathbin{\pm}3.35\times10^{6}$ \\
\cline{2-6}
\multirow{3}{*}{Expansion ratio $r_{\rm exp}$} & $1$ & $0.44\mathbin{\pm}10.02$ & $1.04\mathbin{\pm}0.57$ & $0.53\mathbin{\pm}10.67$ & $16.75\mathbin{\pm}1.72\times10^{5}$ \\
 & $1.5$ & $0.58\mathbin{\pm}37.67$ & $2.40\mathbin{\pm}1.08$ & $0.71\mathbin{\pm}31.71$ & $2778.17\mathbin{\pm}1.83\times10^{6}$ \\
 & $2$ & $0.82\mathbin{\pm}222.97$ & $3.78\mathbin{\pm}1.65$ & $0.93\mathbin{\pm}94.12$ & $1.41\times10^{5}\mathbin{\pm}5.28\times10^{6}$ \\
\cline{2-6}
\multirow{3}{*}{Oversampling $\sigma$} & $1.25$ & $0.54\mathbin{\pm}130.92$ & $2.33\mathbin{\pm}1.90$ & $0.64\mathbin{\pm}57.84$ & $2637.79\mathbin{\pm}3.35\times10^{6}$ \\
 & $1.5$ & $0.54\mathbin{\pm}130.92$ & $2.33\mathbin{\pm}1.90$ & $0.64\mathbin{\pm}57.84$ & $2637.79\mathbin{\pm}3.35\times10^{6}$ \\
 & $2$ & $0.54\mathbin{\pm}130.92$ & $2.33\mathbin{\pm}1.90$ & $0.64\mathbin{\pm}57.84$ & $2637.79\mathbin{\pm}3.35\times10^{6}$ \\
\hline
\end{tabular}
\end{sidewaystable}

\end{document}